\documentclass[usenatbib]{mn2e}
\newcommand{\bib}{\bibitem[\protect\citeauthoryear}
\newcommand{\pa}{\,\rlap{\raise 0.5ex\hbox{$\propto$}}{\lower 1.0ex\hbox{$\sim$}}\,}
\usepackage[dvips]{graphicx}
\usepackage{float}
\usepackage[T1]{fontenc}
\usepackage{ae,aecompl}
\begin{document}
\title[FR\,I radio galaxies with lobes]
{Deep imaging of Fanaroff-Riley Class\,I radio galaxies with lobes}
\author[R.A. Laing et al.]
   {R.A. Laing \thanks{E-mail: rlaing@eso.org}$^{1}$, D. Guidetti$^{1,2}$, 
    A.H. Bridle$^3$, P. Parma$^2$, M. Bondi$^2$\\
    $^1$ European Southern Observatory, Karl-Schwarzschild-Stra\ss e 2, D-85748 
    Garching-bei-M\"unchen, Germany \\
    $^2$ INAF - Istituto di Radioastronomia,
              Via Gobetti 101, I--40129 Bologna, Italy\\ 
    $^3$ National Radio Astronomy Observatory, 520 Edgemont Road, Charlottesville,
    VA 22903-2475, U.S.A.}

\date{Received }
\maketitle

\begin{abstract}
We present deep, high-resolution imaging of the nearby Fanaroff-Riley Class\,I
(FR\,I) radio galaxies NGC\,193, B2\,0206+35, B2\,0755+37 and M\,84 at
frequencies of 4.9 and 1.4\,GHz using new and archival multi-configuration
observations from the Very Large Array. In addition, we describe lower-resolution
observations of B2\,0326+39 and a reanalysis of our published images of
3C\,296. All of these radio galaxies show twin jets and well-defined lobes or
bridges of emission, and we examine the common properties of this class of
source. We show detailed images of total intensity, brightness gradient,
spectral index, degree of polarization and projected magnetic-field
direction. The jet bases are very similar to those in tailed twin-jet sources
and show the characteristics of decelerating, relativistic flows. Except on one
side of M\,84, we find that the jets can be traced at least as far as the ends
of the lobes, where they often form structures which we call ``caps'' with sharp outer
brightness gradients. Continuing, but less well collimated flows back into the
lobes from the caps can often be identified by their relatively flat spectral
indices.  The lobes in these radio galaxies are similar in morphology,
spectral-index distribution and magnetic-field structure to those in more
powerful (FR\,II) sources, but lack hot-spots or other evidence for strong shocks at
the ends of the jets. M\,84 may be an intermediate case between lobed and tailed
sources, in which one jet does not reach the end of its lobe, but
disrupts to form a ``bubble''. 
\end{abstract}

\begin{keywords}
galaxies: jets -- radio continuum:general -- magnetic fields --
polarization
\end{keywords}

\section{Introduction}
\label{Introduction}

Relativistic jets are the primary channel of energy loss from accreting
supermassive black holes in many radio galaxies. They also have a major impact
on their surroundings and act as accelerators of the most energetic photons (and
perhaps hadrons) we observe. The present paper forms part of a study of jet
physics in nearby, low-luminosity radio galaxies, specifically those with FR\,I
morphology \citep{FR74}.  We have developed a sophisticated model of FR\,I jets
as relativistic, symmetrical, axisymmetric flows.  By fitting to deep,
high-resolution radio images in total intensity and linear polarization, we have
determined the three-dimensional variations of velocity, emissivity and
magnetic-field ordering in five sources \citep{LB02a,CL,CLBC,LCBH06}.  We have
shown that FR\,I jets decelerate from relativistic ($\beta = v/c \approx 0.8$)
to sub-relativistic speeds on scales of a few kpc and that they are faster
on-axis than at their edges, as expected if they entrain external material.

The physics of boundary-layer entrainment must depend on the composition and
density of the surrounding medium, and in particular on whether the jets
propagate in direct contact with the intergalactic medium (IGM) or are
surrounded by lobes consisting primarily of tenuous and at least partially
relativistic plasma. Of the five sources we have modelled, three have {\em
plumed} or {\em tailed} outer structures wherein most of the extended emission
appears to lie further from the active nucleus than the narrower jets: 3C\,31
\citep{LB02a,3c31ls}, B2\,1553+24 \citep{CL,Young} and NGC\,315
\citep{CLBC,ngc315ls}. We presume that their jets are in direct contact with the
IGM. On the other hand, 3C\,296 \citep{LCBH06} has two {\em lobes} with
well-defined outer boundaries and a diffuse {\em bridge} of emission around the
jets (at least in projection).  A further source, B2\,0326+39 \citep{CL},
clearly has lobes, but it was unclear from published observations
\citep{Bridle91} whether these lobes surround the inner jets.

Our best-fit model for the jets in the lobed FR\,I source 3C\,296 \citep{LCBH06} 
is unusual in that it shows a very large transverse velocity gradient across the
jet except very close to the nucleus, the
ratio of edge to central velocity in this jet being $\la 0.1$, compared with values
$\approx$0.4 for the jets in three tailed sources (the value for 0326+39 is poorly
determined). It is therefore of interest to examine whether jets in other FR\,I
sources whose lobes entirely surround them appear to decelerate differently with
distance from the nucleus than those in tailed FR\,I sources, or those in sources
whose lobes may not extend all the way back to the nucleus, leaving the
inner jets unshielded. Projection complicates interpretation of individual
sources (the lobes may appear superimposed on the jets even if they are not in
physical contact) and it is not always straightforward to separate jet and lobe
emission, but the differences between 3C\,296 and the rest are large. Modelling
of the jets in a small number of lobed sources (which form the majority of
complete samples of low-luminosity radio galaxies; \citealt{PDF96}) should
therefore be enough to decide whether there are systematic differences between
the two classes.

The primary aim of this paper is to present high-quality radio imaging of four
lobed FR\,I sources whose jets are suitable for modelling by our methods. The
models themselves will be presented elsewhere. 
Our approach to jet modelling requires fitting to high-fidelity, deep
images with linear resolution $\la$0.25\,kpc derived from multi-configuration
observations at 4.9\,GHz (C-band) or 8.5\,GHz
(X-band) with the Very Large Array (VLA) at the National Radio Astronomy
Observatory\footnote{The National Radio Astronomy Observatory is a
facility of the National Science Foundation operated under cooperative agreement
by Associated Universities, Inc.}. In order to correct the linear polarization for the effects of Faraday
rotation (small at these frequencies), we also need to observe at several
frequencies in the range 1.3 -- 1.7\,GHz (L-band). High-quality images of
depolarization, rotation measure and spectral index are useful byproducts.  In
this paper, we describe:
\begin{enumerate}
\item details of the observations, 
\item the source morphologies in total intensity at a range of resolutions, 
\item images of the spectral-index distributions and
\item images of the degree of polarization and the apparent magnetic field
  direction, corrected for Faraday rotation.
\end{enumerate}
We also include high-resolution MERLIN\footnote{MERLIN is a UK National Facility
operated by the University of Manchester at Jodrell Bank Observatory on behalf
of STFC.} imaging for two of the sources.  Faraday rotation and depolarization
are analysed in detail by \citet[and in preparation]{Guidetti11}.

\begin{center}
\begin{table}
\caption{The sources. Col.\,1: name (as used in this paper). Col.\,2: alternative
  names. Col.\,3: redshift. Col.\,4: linear scale (kpc\,arcsec$^{-1}$) for our
  adopted cosmology. Col.\,5: reference for redshift. Col.\,6: reference for
  earlier observations of large-scale radio structure.\label{tab:sources}}
\begin{minipage}{80mm}
\begin{tabular}{llllll}
\hline
Name & Alternative & $z$ & kpc&\multicolumn{2}{c}{Ref}\\
     &             &     &arcsec$^{-1}$&O&R\\
\hline
NGC\,193 & PKS\,0036+03 & 0.0147 & 0.300 &9&4\\
         &UGC\,408 &&&&\\
0206+35 & UGC\,1651 & 0.0377 & 0.748 &3&8\\
        & 4C\,35.03 &&&&\\
0755+37 & NGC\,2484 & 0.0428 & 0.845 &9&1\\
M\,84 & 3C\,272.1 & 0.0035 & 0.073 &10&5\\
      &  NGC\,4374 &&&&\\
3C\,296 & NGC\,5532 & 0.0247 & 0.498 &7&6\\
0326+39 &                & 0.0243 & 0.490 &7&2\\
\hline
\end{tabular}
References: 1 \citet{Bondi00}, 2 \citet{Bridle91}, 3 \citet{Falco99}, 4
\citet{Gia11}, 5 \citet{LB87}, 6 \citet{LCBH06}, 7 \citet{Miller02}, 8 \citet{Morganti87}, 9
\citet{Ogando}, 10 \citet{Trager}.
\end{minipage}
\end{table}
\end{center}

We also present an analysis of the spectrum of 3C\,296 which improves on that
given by \citet{LCBH06} and imaging of the large-scale structure of
B2\,0326+39 to trace its lobe emission closer to the nucleus than in earlier studies.
These data complete the documentation of the large-scale structures and
spectral-index distributions for all of the lobed FR\,I sources whose jets we can
currently model. 

Section \ref{Obs-red} describes the new observations and data reduction, Section
\ref{Images} presents our results for the sources individually, and Section
\ref{discuss} outlines the phenomenology of their jets and larger-scale
emission as a prelude to modelling. Section~\ref{summary} is a brief summary.

We adopt a concordance cosmology with Hubble constant, $H_0$ =
70\,$\rm{km\,s^{-1}\,Mpc^{-1}}$, $\Omega_\Lambda = 0.7$ and $\Omega_M =
0.3$. 

\begin{center}
\begin{table*}
\caption{Journal of VLA observations. Col.\,1: source name. Col.\,2: VLA
  configuration. Col.\,3: Date of observation. Col.\,4: centre frequencies for
  the one or two channels observed (MHz). Col.\,5 bandwidth (MHz). Col.\,6: the
  on-source integration time scaled to an array with all 27 antennas
  operational. Col.\,7: VLA proposal code.}
\begin{center}
\begin{tabular}{lclllcl}
\hline
Source & Config- &    Date     & $\nu$ & $\Delta\nu$  & t & Proposal \\
       & uration &             & [MHz]     & [MHz]      & [min] & code\\
\hline
NGC\,193& A & 2007 Jun 02& 4885.1, 4835.1   & 50 &280& AL693 \\
        & A & 2007 Jun 28& 4885.1, 4835.1   & 50 & 90& AL693 \\
        & A & 2007 Aug 22& 4885.1, 4835.1   & 50 & 41& AL693 \\
        & A & 2007 Aug 23& 4885.1, 4835.1   & 50 & 88& AL693 \\
        & B & 2007 Nov 05& 4885.1, 4835.1   & 50 &318& AL693 \\
        & B & 2007 Nov 16& 4885.1, 4835.1   & 50 &121& AL693 \\
        & C & 2008 May 24& 4885.1, 4835.1   & 50 &223& AL693 \\
        & D & 2007 Mar 11& 4885.1, 4835.1   & 50 & 53& AL693 \\
        & A & 2007 Jun 28& 1365.0           & 25 & 83& AL693 \\
        & A & 2007 Aug 22& 1365.0           & 25 & 39& AL693 \\
        & A & 2007 Aug 23& 1365.0           & 25 & 97& AL693 \\
        & B & 2007 Nov 16& 1365.0           & 25 &148& AL693 \\
        & C & 2008 May 24& 1365.0           & 25 & 61& AL693 \\
0206+35 & A & 2008 Oct 13& 4885.1, 4835.1   & 50 &486& AL797 \\
        & A & 2008 Oct 18& 4885.1, 4835.1   & 50 &401& AL797 \\
        & B & 2003 Nov 17& 4885.1, 4835.1   & 50 &254& AL604\\
        & C & 2004 Mar 20& 4885.1, 4835.1   & 50 &88 & AL604\\
        & A & 2004 Oct 24& 1385.1, 1464.9   & 25 &189& AL604\\
        & B & 2003 Nov 17& 1385.1, 1464.9   & 25 &110& AL604\\
0755+37 & A &2008 Oct 05& 4885.1, 4835.1   & 50 &477& AL797 \\
        & A & 2008 Oct 06& 4885.1, 4835.1   & 50 &383& AL797 \\
        & B & 2003 Nov 15& 4885.1, 4835.1   & 50 & 332& AL604\\
        & B & 2003 Nov 30& 4885.1, 4835.1   & 50 & 169& AL604\\
        & C & 2004 Mar 20& 4885.1, 4835.1   & 50 & 125& AL604\\
        & D & 1992 Aug 2 & 4885.1, 4835.1   & 50 &55  & AM364\\
        & A & 2004 Oct 25& 1385.1, 1464.9   &12.5&450& AL604\\
        & B & 2003 Nov 30& 1385.1, 1464.9   &12.5&160& AL604\\
        & C & 2004 Mar 20& 1385.1, 1464.9   &12.5& 21& AL604\\
M\,84   & A & 1980 Nov 09 & 4885.1           & 50 &223& AL020 \\
        & A & 1988 Nov 23 & 4885.1, 4835.1   & 50 &405& AW228 \\
        & A & 2000 Nov 18 & 4885.1, 4835.1   & 50 &565& AW530 \\
        & B & 1981 Jun 25 & 4885.1           & 50 &156& AL020 \\
        & C & 1981 Nov 17 & 4885.1           & 50 &286& AL020 \\
        & C & 2000 Jun 04 & 4885.1, 4835.1   & 50 &138& AW530 \\
        & A & 1980 Nov 09 & 1413.0           & 25 & 86& AL020 \\
        & B & 1981 Jun 25 & 1413.0           & 25 & 29& AL020 \\
        & B & 2000 Feb 09 & 1385.1, 1464.9   & 50 & 30& AR402 \\
0326+39 & D & 1997 Dec 13 & 4885.1, 4835.1 & 50 & 11& AR386 \\
        & D & 1997 Dec 16 & 4885.1, 4835.1   & 50 & 32& AR386 \\
        & C & 1998 Dec 4  & 1464.9, 1414.9   & 50 & 11& AR386 \\
        & D & 1997 Dec 13 & 1464.9, 1385.1   & 50 &  6& AR386 \\
        & D & 1997 Dec 16 & 1464.9, 1385.1   & 50 &  9& AR386 \\
\hline 
\end{tabular}
\end{center}
\label{tab:journal}
\end{table*}
\end{center}

\section{Observations and data reduction}
\label{Obs-red}

\subsection{The sources}
\label{Sources}

We selected four bright FR\,I sources for which available data suggested: (a)
that full synthesis observations with the VLA would achieve signal-to-noise
sufficient to image the linearly polarized emission from their counter-jets with
several beamwidths resolution transverse to the radio features, as required by
our modelling methods, and (b) that the jets have formed lobe-like structures rather than diffuse outer
plumes.  Three of the sources, NGC\,193, B2\,0206+35 and B2\,0755+37 are
analogous to 3C\,296 in having well-defined outer boundaries. The fourth, M\,84,
is in some respects an intermediate case between the two classes, as we
discuss below. We analysed a combination of new and archival datasets chosen to
give good spatial-frequency coverage at two or three frequencies.

In addition, we improved our low-resolution images of 3C\,296
\citep{LCBH06}. Finally, we analysed shorter, low-resolution, archival VLA
observations for B2\,0326+39.

Alternative names, redshifts, linear scales and references for all of the
sources are given in Table~\ref{tab:sources} (we drop the B2 from source names
from now on). A journal of observations is given in
Table~\ref{tab:journal}\footnote{\citealt{LCBH06} give a full description of the
observations of 3C\,296; this is not repeated here.}.

\begin{center}
\begin{table*}
\caption{Resolutions and noise levels for the images used in this
  paper. Col.\,1: source name. Col.\,2: resolution (FWHM, in arcsec). Col.\,3:
  frequency, in MHz.  Col.\,4: VLA configurations used in the image (not relevant for
  the two MERLIN observations). Cols\,5 and 6: deconvolution method for $I$ and
  $Q/U$ images, respectively (MR: multi-resolution {\sc clean}; CL:
  single-resolution {\sc clean}; ME: maximum entropy). Cols\,7 and 8: off-source
  noise levels. $\sigma_I$ is the noise level on the $I$ image; $\sigma_P$ the
  average of the noise levels for $Q$ and $U$.  All noise levels were determined
  before correction for the primary beam response and larger values are
  appropriate at significant distances from the field centre. Col.\,9: the
  approximate maximum scale of structure imaged reliably \citep{ObsSS}.
  Cols\,10 and 11: flux densities measured from the image and derived from
  single-dish measurements, respectively. The single-dish flux densities were
  taken from the references in Col.\,12. They have been corrected as necessary
  to put them on the standard scale of \citet{Baars} and interpolated to our
  observing frequencies.
\label{tab:images}}
\begin{minipage}{170mm}
\begin{tabular}{llllllrrrlll}
\hline
Source & FWHM       & Freq  & Config- & \multicolumn{2}{c}{Method} &
\multicolumn{2}{c}{rms noise level} & Max &\multicolumn{2}{c}{$I_{\rm int}$/Jy}&Reference \\
       & [arcsec]   & [MHz]   & urations& $I~$&$QU$  &
\multicolumn{2}{c}{[$\mu$Jy\,beam$^{-1}$]}& scale & Image & SD \\
       &                  &        & &&             & $\sigma_I$ & $\sigma_P$ & [arcsec] &\\
\hline
NGC\,193& 0.45 & 4860.1    & ABCD &ME&CL& 8.6 & 8.2 &300&     &     &\\
& 1.35 & 4860.1    & ABCD &MR&CL& 7.1 & 7.4 &300& $0.79\pm 0.02$ &
$0.81\pm 0.04$ &3\\
& 1.60 & 4860.1    & ABCD &MR&CL& 7.5 & 7.6 &300& $0.78\pm 0.02$ &
$0.81\pm 0.04$ &3\\
& 1.60 & 1365.0    & ABC  &MR&MR&  36 &  31 &900& $1.96\pm 0.04$ & 1.84   &7\\
& 4.05 & 4860.1    & ABCD &CL&CL& 10  & 10  &300& $0.78\pm 0.02$ &
$0.81\pm 0.04$ &3\\ 
& 4.05 & 1365.0    & ABC  &MR&MR&  54 &  25 &900& $1.95\pm 0.04$ & 1.84   &7\\ 
0206+35 &0.16& 1658.0     & MERLIN&CL&$-$&41 & $-$ &2.5  &    &\\
 &0.35 & 4860.1    & ABC  &MR&MR& 7.2 & 7.1 &300&     &    &\\
 & 1.20 & 4860.1    &  BC  &MR&MR& 12  &  12 &300& $0.90\pm 0.02$ &
$0.98\pm 0.12$ & 2\\
& 1.20 & 1464.9    & AB   &MR&MR& 25  & 26  &300& $2.12\pm 0.04$ &2.13&5\\
& 1.20 & 1385.1    & AB   &MR&MR& 25  & 26  &300& $2.12\pm 0.04$ &2.22&5\\
& 1.20 & 1425.0    & AB   &MR&$-$& 19  &$-$  &300& $2.12\pm 0.04$ &2.18&6\\
& 4.50 & 4860.1    &  BC  &MR&$-$& 18  &$-$  &300& $0.90\pm 0.02$ &
$0.98\pm 0.12$ & 2\\
& 4.50 & 1425.0    & AB   &MR&$-$& 38  & $-$ &300& $2.13\pm 0.04$ &2.18&6\\
0755+37  &0.14& 1658.0     & MERLIN&CL&$-$&68 & $-$ &2.5  &    &\\
 & 0.40 & 4860.1    & ABCD &MR&MR& 8.0 & 7.1 &300&     &    &\\
 & 1.30 & 4860.1    & BCD  &MR&MR&7.8  &7.9  &300& $1.26\pm 0.03$
&$1.27\pm 0.02$ &4\\
 & 1.30 & 1464.9    & ABC  &MR&MR& 28  & 28  &300&$2.60\pm 0.05$&$2.53\pm 0.09$&4\\
 & 1.30 & 1385.1    & ABC  &MR&MR& 27  & 26  &300&$2.74\pm 0.05$&$2.62\pm 0.09$&4\\
 & 1.30 & 1425.0    & ABC  &MR&$-$& 20  &$-$ &300&$2.64\pm 0.05$&$2.57\pm 0.09$&4\\
 & 4.00 & 4860.1    & BCD &MR&MR&14   & 12  &300& $1.25\pm 0.03$ &
$1.27\pm 0.02$ &4\\
 & 4.00 & 1464.9    & ABC  &MR&MR& 44  &  42 &300&$2.59\pm 0.05$&$2.53\pm 0.09$&4\\
 & 4.00 & 1385.1    & ABC  &MR&MR& 46  &  36 &300&$2.73\pm 0.05$&$2.62\pm 0.09$&4\\
& 4.00 &  1425.0    & ABC  &MR&MR& 32  &$-$  &300&$2.65\pm 0.05$&$2.57\pm 0.09$&4\\
M\,84   & 0.40 & 4860.1\footnote{Although the 1980 and 1981 observations have a
  centre frequency of 4885.1\,MHz, the weighted mean for the combined dataset is
  still close to 4860.1\,MHz}& ABC  &MR&MR& 11  & 10  &300&                &
               &  \\
  & 1.65 & 4860.1$^a$& ABC  &MR&MR& 15  & 11  &300& $2.94\pm 0.06$ &
$2.88\pm 0.08$ &5\\
  & 1.65 & 1413.0    & AB   &MR&MR& 140 & 140 &120& $6.03\pm 0.12$ &
$6.44\pm 0.24$ &5\\
  & 4.5  & 4860.1$^a$& C    &MR&MR&  23 & 20  &300& $2.98\pm 0.06$ &
$2.88\pm 0.08$ &5\\
 & 4.5  & 1464.9    &  B   &MR&MR& 120 & 70   &120& $6.14\pm 0.12$ &
$6.32\pm 0.24$ &5\\
  & 4.5  & 1413.0    & AB   &MR&MR& 210 &150  &120& $5.99\pm 0.12$ &
$6.44\pm 0.24$ &5\\
  & 4.5  & 1385.1    &  B   &MR&MR& 130 & 69  &120& $6.46\pm 0.13$ &
$6.51\pm 0.24$ &5\\
3C\,296    & 5.5&8460.1&ABCD&MR&$-$&14  &$-$& 180 & $1.34\pm 0.03$ &$1.20\pm 0.10$&5\\
           & 5.5&1479.0&BCD&MR&$-$&24   &$-$& 900 & $4.10\pm 0.08$ &$4.08\pm 0.18$&5\\
0326+39&18.0&4860.1& D&CL&$-$& 47   &$-$& 300 & $0.60\pm 0.01$ &$0.62\pm 0.09$
&1 \\
           &18.0&1425.0&CD&CL&$-$&140   &$-$& 900 & $1.47\pm 0.03$ &$1.30$        &6 \\
\hline
\end{tabular}
References: 1 \citet{BWE}, 2 \citet{87GB}, 3 \citet{PMN}, 4 \citet{Kuehr81}, 5
\citet{LP}, 6 \citet{WB}, 7 \citet{PKS}.
\end{minipage}
\end{table*}
\end{center}

\subsection{VLA data reduction}
\label{Reduction}

The VLA data listed in Table~\ref{tab:journal} were calibrated and imaged using
the {\sc aips} software package, following standard procedures with a few
additions. The flux-density scale was set using observations of 3C\,286 or
3C\,48 and (except for 0326+39) the zero-point of ${\bf E}$-vector position
angle was determined using 3C\,286 or 3C\,138, after calibration of the
instrumental leakage terms.  The main deviations from standard methods were as
follows.

Firstly, we used the routine {\sc blcal} to compute closure corrections for the
4.9-GHz observations. This was required to correct for large closure errors on
baselines between EVLA and VLA antennas in observations from 2007 onwards 
\citep{webref}, but also improved a number of the earlier datasets. Whenever
possible, we included observations of the bright, unresolved calibrator 3C\,84
for this purpose; if it was not accessible during a particular observing run, we used
3C\,286. We found that it was  not adequate to use the standard
calibration (which averages over scans) to compute the baseline corrections, as
phase jumps during a calibrator scan caused serious errors in the derived
corrections. We therefore self-calibrated the observations in amplitude and
phase with a solution interval of 10\,s before running {\sc blcal}. We assumed a
point-source model for 3C\,84 and the well-determined {\sc clean} model supplied
with the {\sc aips} distribution for 3C\,286.

Secondly, we imaged in multiple facets to cover the inner part of the primary
beam at L-band and to image confusing sources at large distances from the phase
centre in all bands.  Before combining configurations, we subtracted in the
$(u,v)$ plane all sources outside a fixed central field. For 0755+37 at L-band,
this procedure failed to remove sidelobes at the centre of the field from a
bright confusing source close to the half-power point of the primary beam. The
reason is that the VLA primary beam is not azimuthally symmetric, so the
effective complex gain for a distant source is not the same as that at the
pointing centre and varies with time in a different way. We used the {\sc aips}
procedure {\sc peelr} to remove the offending source from the $(u,v)$ data for
each configuration before combining them.

Finally, we corrected for variations in core flux density and amplitude scale
between observations as described in \citet{LCBH06}.

J2000 coordinates are used throughout this paper.  If positions from archival
data were originally in the B1950 system, then (u,v,w) coordinates were
recalculated for J2000 before imaging. The astrometry for each of the sources
was set using the A-configuration observations, referenced to a nearby phase
calibrator in the usual manner. Thereafter, the position of the compact core was
held constant during the process of array combination.

The observations of M\,84 in 1980 and 1981 used an earlier and less accurate
value of the position of the phase calibrator B1236+077 (alias J1239+075) than
that currently given in the VLA calibrator manual. We have updated the
astrometry to reflect the improved calibrator position.  The archival L-band
observations of M\,84 taken in 2000 used a pointing centre displaced by
$\approx$1.1\,arcmin from the centre of the source.

The C-band data were usually taken in two adjacent 50-MHz frequency channels,
which were imaged together. The L-band channels were also imaged together in
$I$, in order to derive spectral-index images.  For all sources except
0326+39, they were also imaged independently, primarily for analysis of
linear polarization.

In order to avoid the well-known problems introduced by the conventional {\sc
clean} algorithm for well-resolved, diffuse brightness distributions,
total-intensity images at the higher resolutions were produced using the
multi-scale {\sc clean} algorithm as implemented in the {\sc aips} package
\citep{MSC} or, in one case, a maximum-entropy algorithm \citep[used as
described by \citealt{LP91}]{CE}.  The standard single-resolution {\sc clean} was
found to be adequate for the lowest-resolution $I$ images.  Stokes $Q$ and $U$
images were {\sc clean}ed using one or more resolutions (we found few
differences between single and multiple-resolution {\sc clean} for these images,
which have little power on large spatial scales). All of the images were
corrected for the effects of the antenna primary beam.

In general, the deep 4.9-GHz images have off-source noise levels very close to
those expected from thermal noise in the receivers alone. There are a few faint
artefacts on the $I$ images. These are visible as concentric rings around the
bright cores and, for NGC\,193 at 1.35 and 1.6-arcsec resolution, a quadrupolar
pattern at the $2\sigma$ level. These are due to errors in cross-calibration of
the different array configurations, which were particularly troublesome due to
the low declination of this source.  The integrations for all of the L-band
images and for the C-band image of 0326+39 are shorter and confusion from
sources outside the field of view is worse, so noise levels are correspondingly
higher.

Finally, we produced improved $I$ images from self-calibrated L- and X-band
visibility data for 3C\,296 \citep{LCBH06} using the multi-resolution {\sc
clean} algorithm.

As a check on the amplitude calibration and imaging of the $I$ images used in
spectral-index analysis, we integrated the flux densities using the {\sc
aips} verb {\sc tvstat}. We estimate that our errors are dominated by a residual
scale error of $\approx$2\%. All of the results are in excellent agreement with
single-dish measurements (Table~\ref{tab:images}).
 
The configurations, resolutions, deconvolution algorithms and noise levels for
the final images are listed in Table~\ref{tab:images}.  The noise levels were
measured before correction for the primary beam, and are appropriate for the
centre of the field.

\begin{figure*}
\includegraphics[width=18cm]{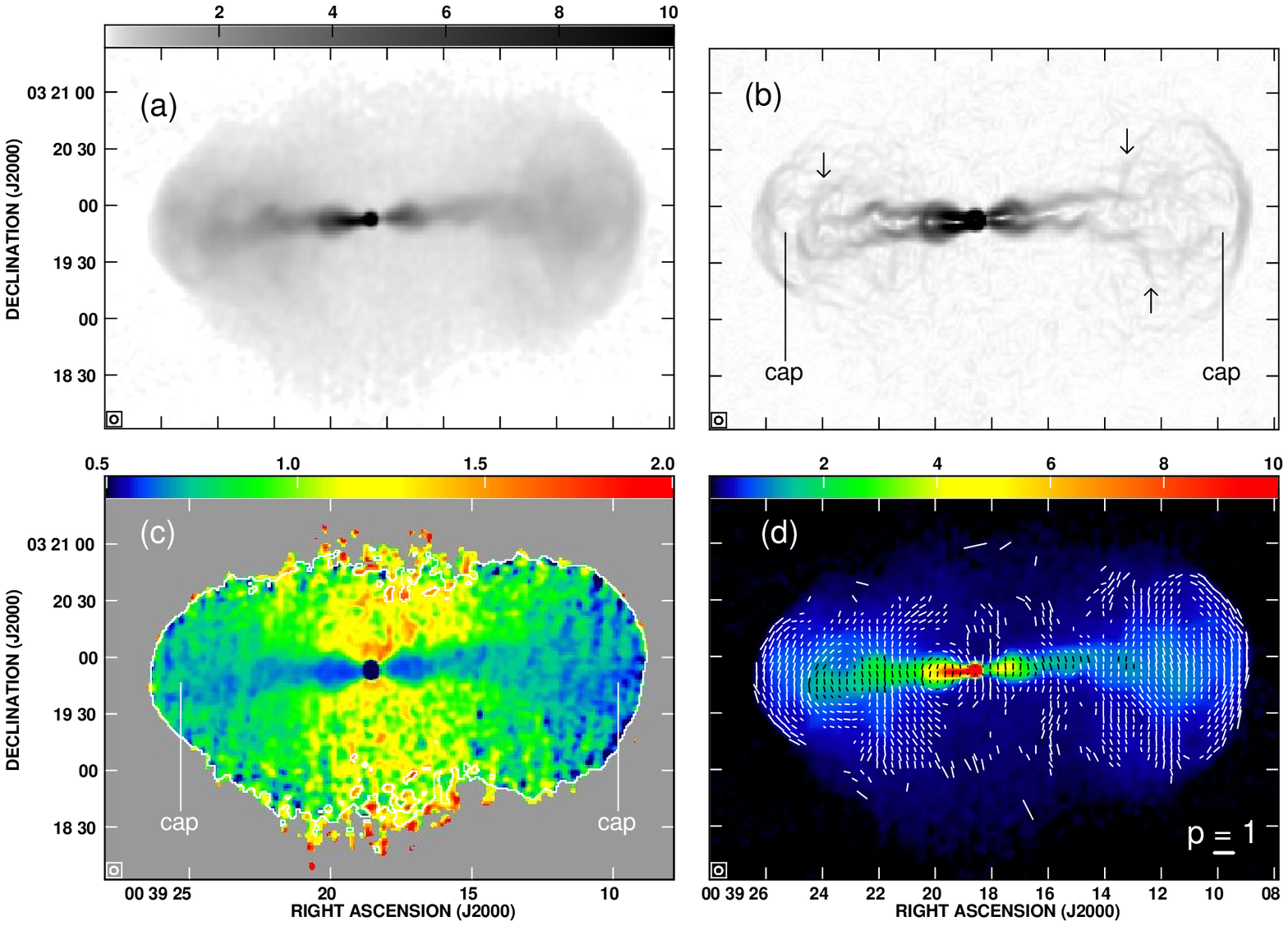}
\caption{Images of NGC\,193 at 4.05-arcsec resolution. (a) Total intensity at
  4.9\,GHz. (b) Intensity gradient for the image in panel (a). The arrows mark
  the high brightness gradients which coincide with the edges of the
  flat-spectrum caps (which are also marked). (c) Spectral-index image between
  4.9 and 1.4\,GHz. Values outside the white contour are lower limits. The caps
  are indicated. (d) Vectors with directions along ${\bf B}_{\rm a}$ and
  magnitudes proportional to $p_{4.9}$, plotted on a false-colour
  image of total intensity at 4.9\,GHz.
\label{fig:ngc193all}}
\end{figure*}

\begin{figure*}
\includegraphics[width=11.75cm]{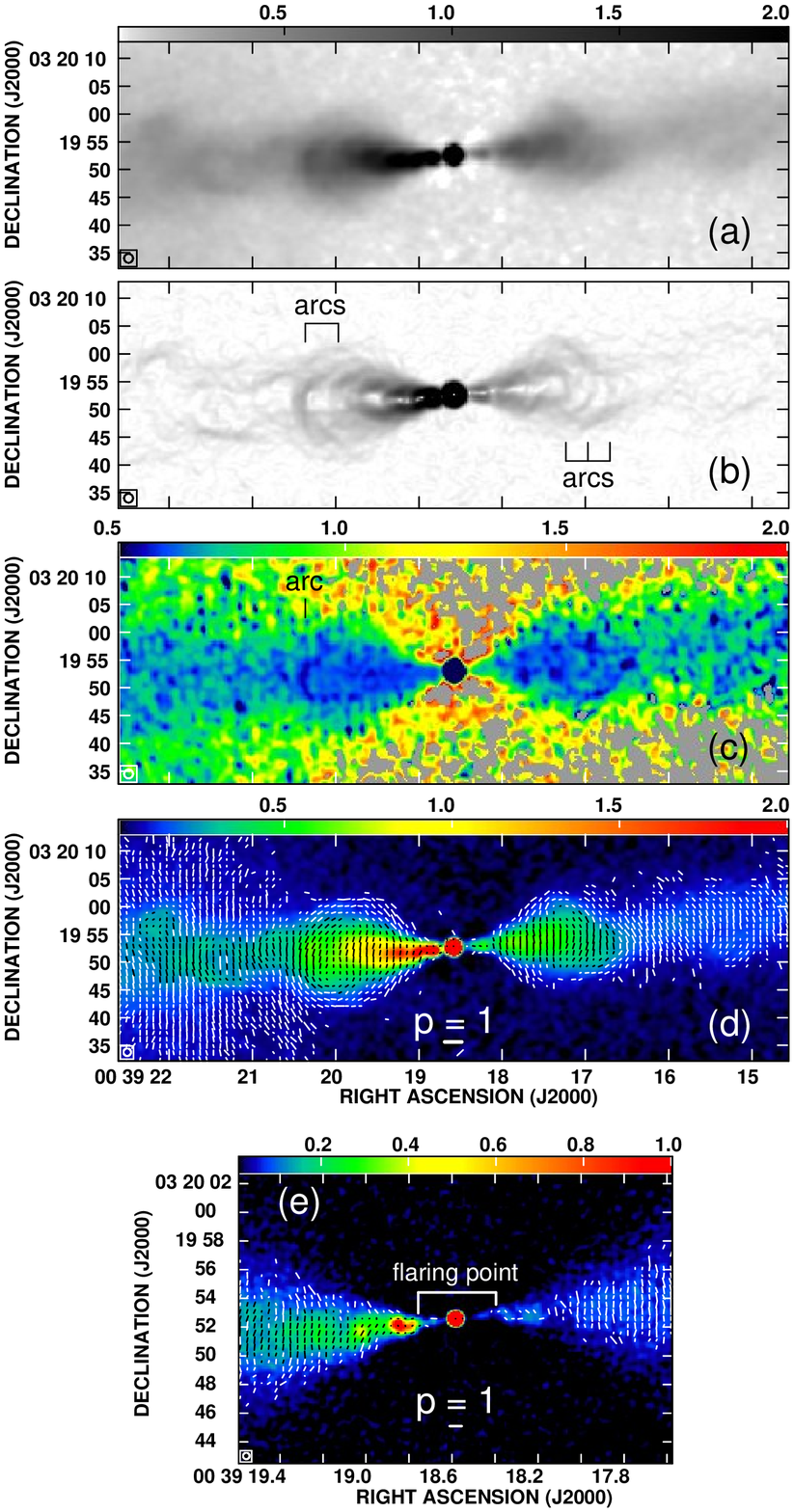}
\caption{High-resolution images of the inner jets of NGC\,193.
  (a) Total intensity at 4.9\,GHz with 1.6\,arcsec FWHM resolution. (b) Brightness
  gradient derived from the image in (a). Sharp steps in brightness (``arcs'')
  are indicated. (c) Spectral index between 4.9 and 1.4\,GHz at 1.6-arcsec
  resolution. The most prominent of the arcs, which shows a flattening in
  spectral index, is marked. (d) Vectors with directions along ${\bf B}_{\rm a}$
  and magnitudes proportional to $p_{4.9}$, plotted on false-colour images of
  total intensity at 4.9\,GHz with 1.35\,arcsec FWHM resolution.  The
  vectors have been corrected for Faraday rotation using a 4.05-arcsec FWHM
  resolution RM image.
  (e) As in panel (d), but for the innermost jet regions at 0.45-arcsec resolution.
\label{fig:ngc193hires}}
\end{figure*}

\subsection{MERLIN observations and reduction}
\label{MERLIN}

We also present MERLIN imaging in total intensity only for two of the sources.
0206+35 was observed for a total time of about 14 hours. The array included the
following telescopes: Defford, Cambridge, Knockin, Wardle, Darnhall, Mk2, Lovell
and Tabley. The observations were carried out at 1420 MHz with a bandwidth of 15
MHz, in each of left and right circular polarizations.  The nearby compact
source 0201+365 was used as the phase calibrator and the flux-density scale was
determined using 3C\,286. The data were edited, corrected for
elevation-dependent effects and non-closing errors and flux-calibrated using the
standard MERLIN analysis programs. Imaging and self-calibration were again
performed using the {\sc aips} package. The off-source image rms after
self-calibration was close to that expected from receiver noise alone. The
MERLIN observations of 0755+37 were described by \citet{Bondi00}.

The parameters of both MERLIN images are given in Table~\ref{tab:images}.

\section{Images}
\label{Images}

\subsection{General}

\begin{figure*}
\includegraphics[width=13.25cm]{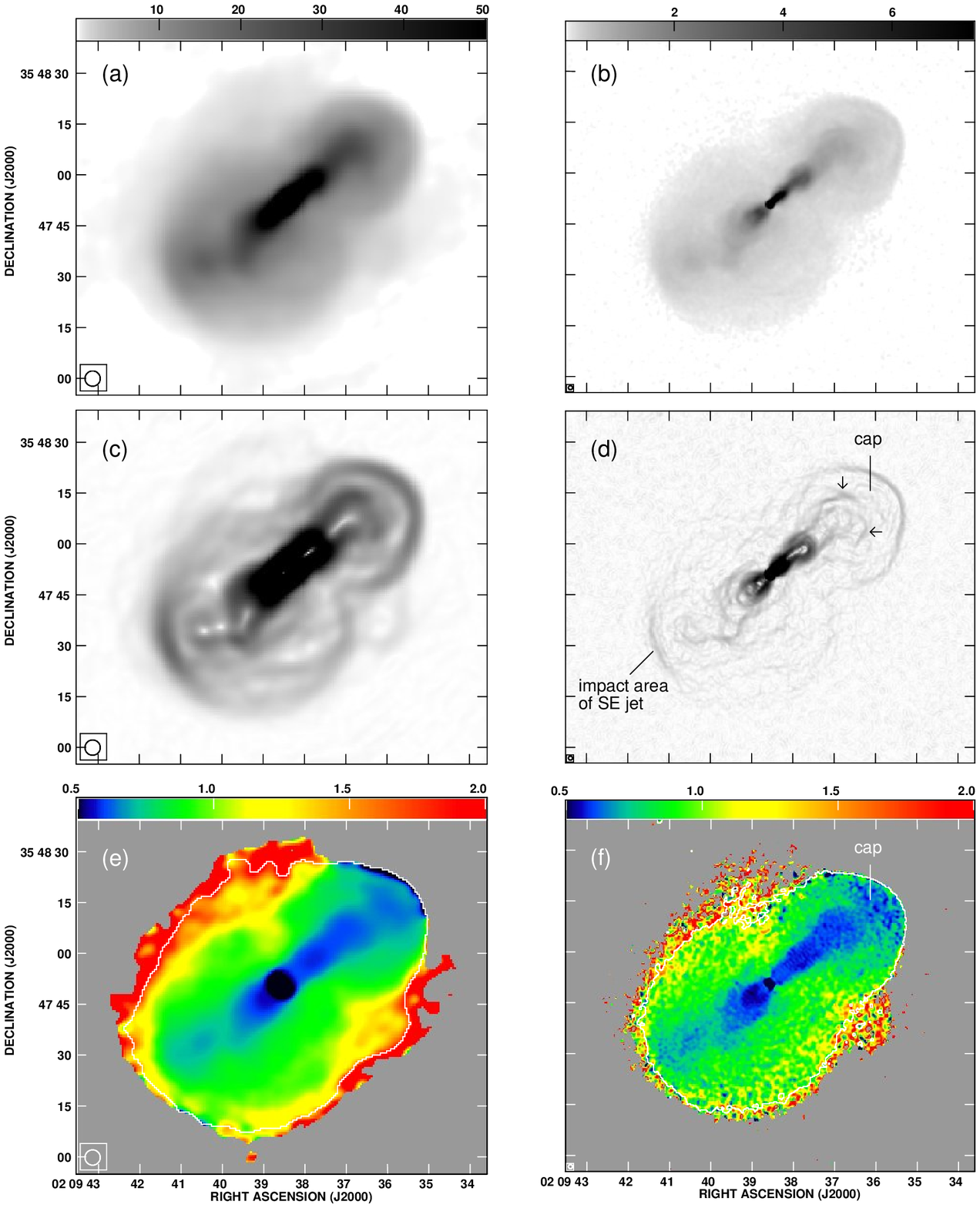}
\caption{Images of 0206+35. (a) Total intensity at 1.4\,GHz with 4.5-arcsec
  FWHM resolution. (b) Total intensity at 4.9\,GHz, 1.2-arcsec FWHM. (c) Intensity gradient
  at 1.4\,GHz, 4.5-arcsec FWHM. (d) Intensity gradient at 1.4\,GHz, 1.2-arcsec
  FWHM. The arrows mark the high brightness gradients around the inner boundary
  of the flat-spectrum cap. (e) Spectral index between 4.9 and 1.4\,GHz,
  4.5-arcsec FWHM. (f) Spectral index between 4.9 and 1.4\,GHz, 1.2-arcsec
  FWHM. In panels (e) and (f), values outside the contours are lower limits.
\label{fig:0206all}}
\end{figure*}

\begin{figure*}
\includegraphics[width=10cm]{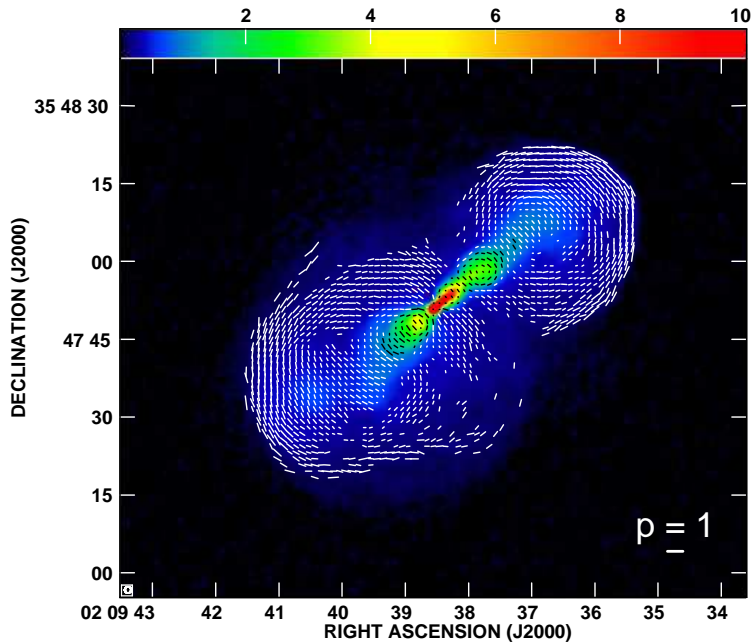}
\caption{Vectors with directions along the apparent magnetic field and lengths
  proportional to the degree of polarization at 4.9\,GHz, superimposed on a
  false-colour image of total intensity across 0206+35 at the same
  frequency. The resolution is 1.2\,arcsec FWHM. The apparent field directions
  were derived using a three-frequency rotation-measure fit \citep{Guidetti11}.
\label{fig:0206vec}}
\end{figure*}

Our conventions for Figs~\ref{fig:ngc193all} -- \ref{fig:0326_i_si18} and the
descriptions in the text are as follows.
\begin{enumerate}
\item Images of total intensity, $I$, are shown as grey-scales, over ranges indicated
  by the labelled wedges. The units are mJy\,beam$^{-1}$.
\item We also show grey-scales of intensity gradient, $|\nabla I|$, approximated 
  using a Sobel filter \citep{Sobel}.
\item We use the notation $P = (Q^2+U^2)^{1/2}$ for polarized intensity and $p =
  P/I$ for the degree of linear polarization.  $p_\nu$ is the degree of
  polarization at frequency $\nu$ (in GHz).  All values of $P$ have been
  corrected for Ricean bias \citep{WK}. Linear polarization is illustrated by
  plots in which vectors with lengths proportional to the degree of polarization
  at 4.9\,GHz ($p_{4.9}$) and directions along the apparent magnetic field
  (${\bf B}_{\rm a}$) are superposed on false-colour images of either $I$ (again
  with a labelled wedge indicating the range) or $|\nabla I|$.  A value of $p =
  1$ is indicated by the labelled bar.  The apparent field direction is $\chi_0
  + \pi/2$, where $\chi_0$ is the ${\bf E}$-vector position angle corrected to
  zero-wavelength by fitting to the relation $\chi(\lambda^2) = \chi_0 + {\rm
  RM}\lambda^2$ for foreground Faraday rotation derived from the images in
  Table~\ref{tab:images} (RM is the rotation measure).  In some sources, we used
  RM images at lower resolution to correct the position angles, as detailed in
  the captions. This procedure is valid if the RM varies smoothly over the
  low-resolution image, and maximises the area over which we can determine the
  direction of the apparent field.  Vectors are plotted where: (a) $I \geq 5
  \sigma_I$, (b) $P \geq 3 \sigma_P$ (the noise levels are given in
  Table~\ref{tab:images}) and (c) the RM is well-determined. The RM images for
  0206+35 and M\,84 were determined using three and four frequencies,
  respectively and are shown in \citet{Guidetti11}; that for 0755+37 (also from
  three frequencies) will be described by Guidetti et al. (in preparation). For
  NGC\,193, images were only available at 4.86 and 1.365\,GHz. The integrated
  RM of the source is small ($-18 \pm 2$\,rad\,m$^{-2}$; \citealt*{SKB}) as are
  the variations of position-angle difference across the source. We are
  therefore confident that a two-frequency RM determination is adequate in this
  case (the correction required to derive ${\bf B}_{\rm a}$ is in any case very small).
\item Spectral index, $\alpha$, is defined in the sense $S(\nu) \propto
\nu^{-\alpha}$ and we use $\alpha^{\nu_2}_{\nu_1}$ for the index between
frequencies ${\nu_1}$ and ${\nu_2}$ (in GHz). In the false-colour images of
spectral index, the input $I$ images at the lower frequency are always blanked
for $I < 3\sigma_I$.\footnote{The spectral-index image for M\,84 has additional
blanking, as noted in the caption of Fig.~\ref{fig:m84all}(c)} In cases where
the areas over which emission was detected were essentially the same at both
frequencies, we also blanked the higher-frequency image at $3\sigma_I$. If
significant areas were detected only at the lower frequency, we did not blank
the higher-frequency image. Instead, we plot a single contour which indicates the
boundary of the region where the source is detected at $I \geq 3\sigma_I$ at the
higher frequency. Outside this contour, the spectral indices are lower limits.
We have carefully inspected the spectral-index images to check for edge effects
and zero-level problems. We are confident that the values and lower limits in
all of the unblanked regions are reliable except where explicitly stated and
that the steep spectra seen at the edges of the  lobes in several sources are
real.  
\item The restoring beam (FWHM) is shown at the bottom of each plot.
\item We refer to parts of the sources by the abbreviations N, S, E, W (for
  North, South, East, West) etc.
\item We refer to the {\em main} (brighter) and {\em counter} (fainter) jets.  
\end {enumerate}

\subsection{NGC\,193}
\label{ngc193}

Fig.~\ref{fig:ngc193all}(a) shows the total-intensity distribution over NGC\,193
at 4.9 GHz and 4.05-arcsec FWHM resolution. The symmetrical jets appear to
broaden rapidly and also bend away from their initial straight path as they
reach the midpoints of symmetric lobes.  The lobes both have well-defined 
leading edges that approximate arcs of circles in projection on the sky (the W lobe 
having a larger radius of curvature), but they lack hot spots that might mark the 
termination points of the jets.  A broad, faint emission bridge fills the central 
region of the source
between the lobes and appears to be wider than the lobes in the N-S
direction in the centre of the source (similar to the ``wings'' observed in some
FR\,II sources; \citealt[and references therein]{wings}).  Figs~\ref{fig:ngc193all}(b) 
and (c) taken together show that broad ``caps'' of emission can be delineated
in both lobes by enhanced intensity gradients and by lower-than-average 
values of $\alpha^{4.9}_{1.4}$. The intensity
gradients are largest at the outer edges of these caps, but there are also gradient
features within the lobes (marked with arrows on Fig.~\ref{fig:ngc193all}b)
which coincide with the edges of the flatter-spectrum region. The regions where
the jets are most prominent have low spectral indices around 0.6. The spectral
index at the trailing edges of the caps steepens smoothly to $\alpha \approx
0.9$ where the emission merges with the broader, symmetric lobes.  The most
diffuse lobe emission has spectral indices increasing from $\approx$1 at the edges
of the most elongated parts of the lobes to $\approx$1.4 near the centre of the
source, a spectral-index pattern characteristic of lobed radio sources of both
FR classes (see Section~\ref{lobes}).  There are regions of emission with
spectral index approaching 2 at the edges of the faintest emission on the N
and S edges of the source.\footnote{The spectral index within a few arcsec
N and S of the core is affected by artefacts in the 4.9-GHz image.}
Fig.~\ref{fig:ngc193all}(d) shows that the distribution of the apparent magnetic
field direction over the lobes is basically circumferential, while the magnetic
field in the jets is perpendicular to the jets over most of their lengths.  The
structure of the apparent magnetic field in both the jets and the lobes appears
regular, and characteristic of that commonly found in jets of FR\,I sources and
in the lobes of both FR classes.

Fig.~\ref{fig:ngc193hires}(a) shows the total-intensity distribution over the
jets at 4.9\,GHz with a resolution of 1.6\,arcsec FWHM.  Both jets are fairly straight and
similar in overall appearance, exhibiting rapid lateral expansion just beyond
the distance from the unresolved nuclear radio source at which the E jet is
markedly brighter than the W jet.  Their edges are well delineated by steep
transverse intensity gradients near the midpoint of the source
(Fig.~\ref{fig:ngc193hires}b).  The surface brightnesses of both jets decrease
smoothly with distance from the nucleus except at a distance of
$\approx$25\,arcsec, where there are more sudden drops, mostly clearly visible
as ``arcs'' crossing the jets on the gradient image (indicated on
Fig.~\ref{fig:ngc193hires}b).  The overall spectral index of the jet-dominated
emission appears to steepen with distance from the nucleus, but this is almost certainly the result of
superposition of dimming jets on steeper-spectrum lobe emission. The prominent arc in the
E jet is associated with a slight flattening in the spectrum
(Fig.~\ref{fig:ngc193hires}c).

Fig.~\ref{fig:ngc193hires}(d) shows the intensity and apparent magnetic field
distributions over the inner $\approx$45-arcsec regions of both jets at 1.35-arcsec
resolution.  The magnetic field organization over both jets is quite regular,
with the field closest to the jet axis predominantly orientated perpendicular to
the axis. The apparent field at the edges of the inner jets is parallel to the
rapidly-expanding outer isophotes. Farther from the nucleus, the edge field
directions in both jets converge towards the axis to form almost circular
patterns.

Fig.~\ref{fig:ngc193hires}(e) shows the 4.9-GHz total intensity and apparent
magnetic-field distributions over the inner 15 arcsec ($\approx$4.5 kpc) of the
jets at 0.45-arcsec resolution.  Both jets exhibit faint inner regions in the
first $\approx$2 arcsec from the nucleus before they brighten and subsequently
flare.  There is bright, non-axisymmetric knot structure in the first $\approx$4
arcsec of the E jet, downstream of the flaring point \citep{LPdRF}, where
both jets brighten abruptly.

\subsection{0206+35}

\begin{figure}
\includegraphics[width=7cm]{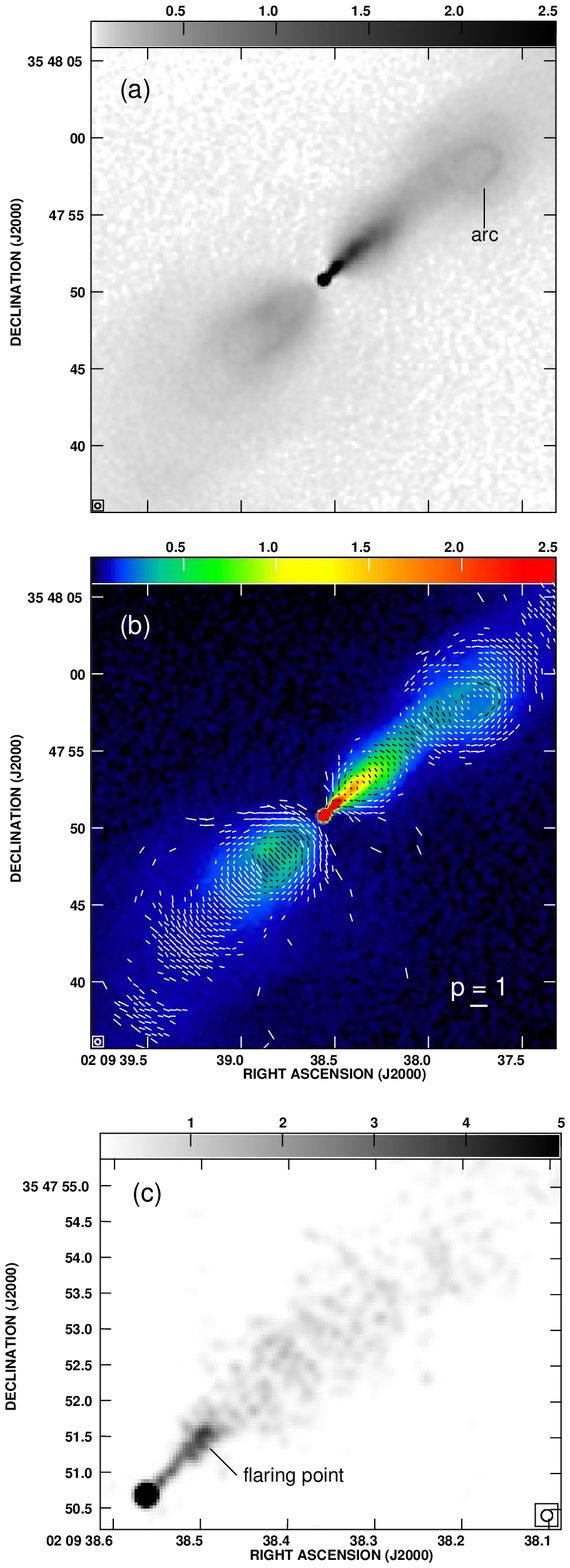}
  \caption{(a) Grey-scale of the 4.9-GHz total-intensity distribution over the
 jets in 0206+35 at 0.35-arcsec FWHM resolution.  The grey-scale range is 0
 -- 2.5\,mJy beam$^{-1}$.  (b) Vectors with lengths proportional to the degree
 of polarization at 4.9\,GHz and directions along the apparent magnetic field,
 superimposed on a false-colour display of the total intensity at 4.9\,GHz.  The
 resolution is 0.35\,arcsec FWHM.  The vector directions are derived from 
 3-frequency RM fits at 1.2-arcsec resolution. (c) Grey-scale of the 1.6-GHz total
 intensity distribution over the inner 5 arcsec of the NW jet and the
 unresolved nuclear source in 0206+35 at 0.16-arcsec FWHM resolution, from
 MERLIN.  The grey-scale range is 0 -- 5\,mJy beam$^{-1}$.
\label{fig:0206hires}}
\end{figure}   

Fig.~\ref{fig:0206all} shows the total-intensity, brightness gradient, and
spectral-index distributions over the whole of 0206+35 at two resolutions.
Fig.~\ref{fig:0206all}(a) at 1.4 GHz, 4.5-arcsec FWHM resolution, shows that the
large scale structure consists of two lobes, overlapping and circular in
cross-section, with well-defined outer edges to the NW and SE of the source,
superimposed on fainter diffuse emission to the N and S.
Fig.~\ref{fig:0206all}(b), at 4.9\,GHz and 1.2-arcsec resolution, shows the
source in more detail but with reduced sensitivity to the largest-scale
emission.  At this resolution, both lobes show sharp outer boundaries.  The
roughly circular edge of the NW lobe protrudes beyond the diffuse emission,
whereas the corresponding feature in the SE is well inside the outer boundary of
the source and is most obvious in the E of the lobe, close to the termination of
the jet.  If the orientation of $\theta \approx 40^\circ$ determined for the
inner jets (Laing \& Bridle, in preparation) also applies to the lobes, then
they are presumably ellipsoidal with an axial ratio $\approx$1.6.
Fig.~\ref{fig:0206all}(b) also shows some internal structure in both jets.  The
NW jet has the brighter base, and both bends and brightens as it enters its
lobe, after which its path meanders. The SE counter-jet appears to expand more
rapidly initially, then also meanders as it enters its lobe.

Figs~\ref{fig:0206all}(c) and (d) show the 1.4-GHz intensity gradient images of
0206+35 at 4.5-arcsec and 1.2-arcsec resolution, respectively.  The edges of
the jets are clearly marked by enhanced intensity gradients at both resolutions,
while significant internal structure is also apparent in the lobes.  Both lobes
exhibit strong brightness gradients at their outer edges in these displays,
corresponding to the sharp boundaries noted earlier.  There is a particularly
striking correlation between the main features of these intensity-gradient
images and of the two 1.4 to 4.9-GHz spectral-index images shown as
Figs.~\ref{fig:0206all}(e) and (f).  The emission inside the brightest intensity
gradients around the jet has a much lower spectral index, typically $<$0.65,
than the $\approx$0.7 to 1.0 spectral index that is prevalent over the rest of the
two spherical lobes.  The diffuse emission outside the lobes has spectral
indices ranging from $\approx$1.05 to $>$2, generally increasing with distance from
the lobes towards the outer edge of the source.  Also notable are the ``fans''
of lower-spectral-index emission that can be traced from the ends of the jets to
the regions at the edges of both lobes that show the most pronounced brightness
gradients.  This is particularly striking in the NW lobe, where a
cap of lower-spectral-index emission is bounded by the high brightness
gradients marked by arrows on Fig.~\ref{fig:0206all}(d) and by the edge of the
lobe. This suggests that the jet outflow has reached the end of the lobe in a
less-collimated, but still identifiable, form.  In the SE lobe, the jet bends to
the N before appearing to impact the edge of the lobe at another enhanced
brightness gradient, marked in Fig.~\ref{fig:0206all}(d).

Fig.~\ref{fig:0206vec} shows that the magnetic field configuration in both
lobes is well ordered and basically circumferential, while the magnetic field in
the jets is predominantly perpendicular to their axis near the centre lines of
the jets, with evidence for parallel field at the jet edges.  The southern edge
of the source is strongly polarized with field tangential to the boundary.

At 0.35-arcsec FHWM resolution (Fig.~\ref{fig:0206hires}a) the lobe emission is
substantially resolved out so the images are dominated by the jets.  The bright
base of the main (NW) jet is clearly centre-brightened, while the corresponding
segment of the counter-jet appears centre-darkened.  While at first sight the
main jet appears to expand more slowly than the counter-jet, {\it the geometry of the
centre-darkened segment of the counter-jet is very similar to that of the main
jet over the first $\approx$10 arcsec}. This suggests a two-component view of
these jets wherein a narrow inner structure, brighter to the NW and fainter to
the SE, is seen superposed on a broader expanding structure that is slightly
brighter to the SE than to the NW.  We will interpret this elsewhere 
(Laing \& Bridle, in preparation) as evidence
for a symmetrical relativistic outflow surrounded by a mildly relativistic
backflow in this source.

The magnetic field is clearly perpendicular to the jet axis over most of the
length of both jets (Fig.~\ref{fig:0206hires}b), but the first few arcsec of the
main jet, where it is brightest, have the magnetic field parallel to the jet
axis.  There is also evidence for oblique, or parallel, magnetic field at the
edges of both jets.

Fig.~\ref{fig:0206hires}(c) shows the bright, narrow base of the NW jet
from MERLIN data at 1.6 GHz.  This higher-resolution (0.16\,arcsec FWHM) image
unambiguously identifies the flaring point in the jet, 0.7\,arcsec from the
nucleus, where it brightens abruptly. This is an important fiducial distance for
our modelling.  The image also shows the start of rapid expansion downstream of
the flaring point.
 
\subsection{0755+37}
\label{0755}

\begin{figure*}
\includegraphics[width=14cm]{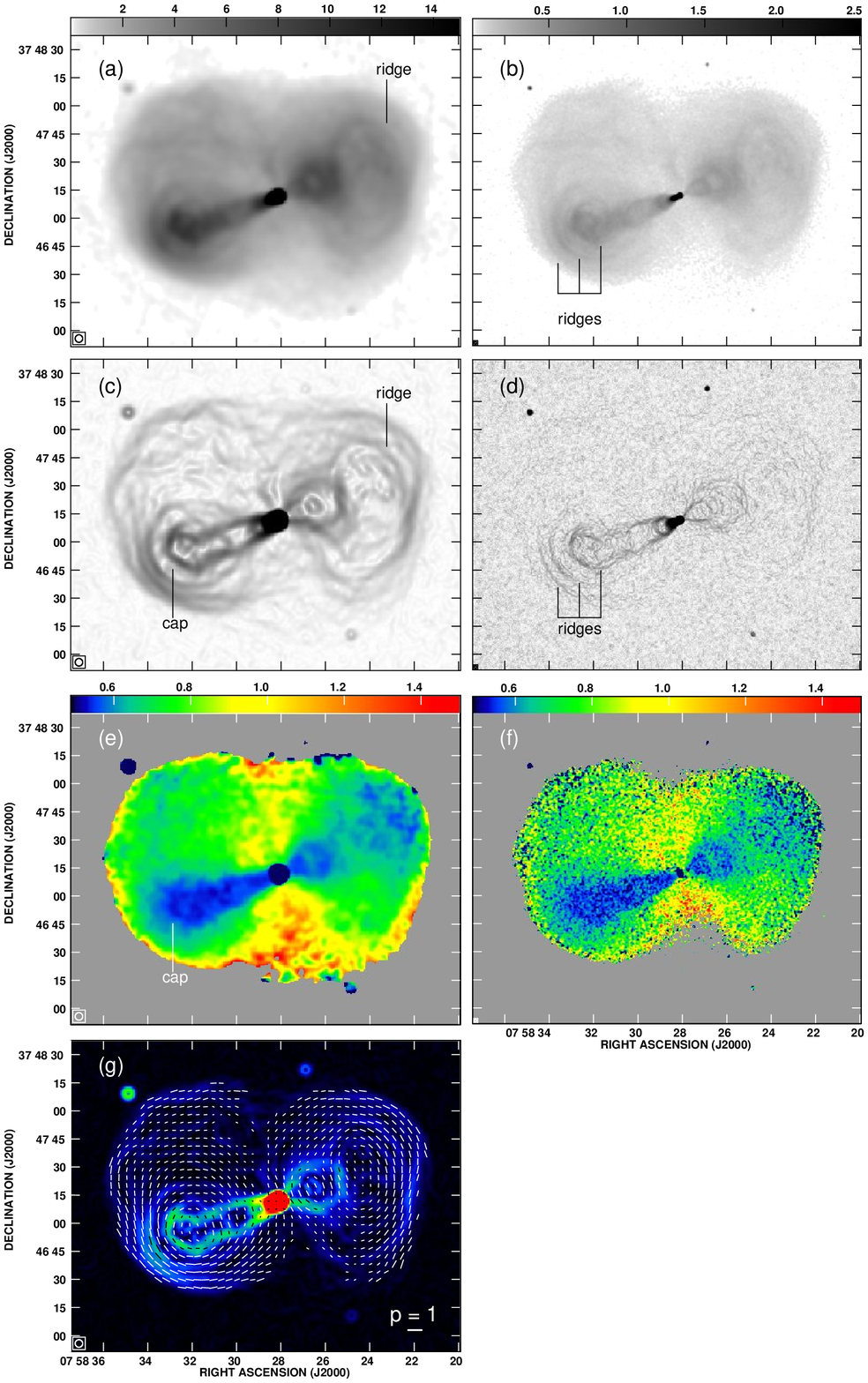}
\caption{Images of the whole of 0755+37 at resolutions of 4.0\,arcsec FWHM
  (panels a, c, e and g) and 1.3\,arcsec FWHM (panels b, d and f). (a) Total
  intensity at 1.4\,GHz. (b) Total intensity at 4.9\,GHz. (c) Intensity gradient
  at 1.4\,GHz. (d) Intensity gradient at 4.9\,GHz. (e) and (f) spectral index
  between 4.9 and 1.4\,GHz. (g) vectors with lengths proportional to $p_{4.9}$
  and directions along the apparent magnetic field from a three-frequency
  rotation-measure fit (Guidetti et al., in preparation), superimposed on the
  intensity gradient at 4.9\,GHz.
\label{fig:0755all}}
\end{figure*}   

Fig.~\ref{fig:0755all} shows the total-intensity, $|\nabla I|$ and $\alpha$
distributions over all of 0755+37 at two resolutions and the polarization at low
resolution. Fig.~\ref{fig:0755all}(a) at 1.4 GHz, 4.0-arcsec FWHM resolution, shows
that the large-scale structure consists of two lobes, again roughly
circular in projection, with well-defined but not particularly sharp outer edges
to the W and E, plus fainter diffuse emission to the N and S.  The E lobe has
a series of narrow ridges and brightness steps, all roughly arcs of circles in
projection, in the region where the brighter jet appears to terminate. They are
recessed from the E boundary of this lobe and some may be the edges of thin
shells.  The structure of the W lobe is unusual, containing some arc-like
features and other structure suggestive of a rapidly decollimating counter-jet
W of the nucleus, as previously described by \citet{Bondi00}, with a
``hole'', or deficit of emission in the region where the counter-jet might be
expected to terminate.  Fig.~\ref{fig:0755all}(b) at 4.9 GHz, 1.3-arcsec FWHM 
resolution, is insensitive to the largest scales of emission to the N and S of
the main source, but clearly shows internal structure in the E lobe, including
the concentric semicircular ridges at the end of the jet (labelled on the
figure).  All of the substructure in the W lobe appears to be resolving out,
though vestiges of the ridge apparent at lower resolution in
Fig.~\ref{fig:0755all}(a) remain.

\begin{figure*}
\includegraphics[width=13cm]{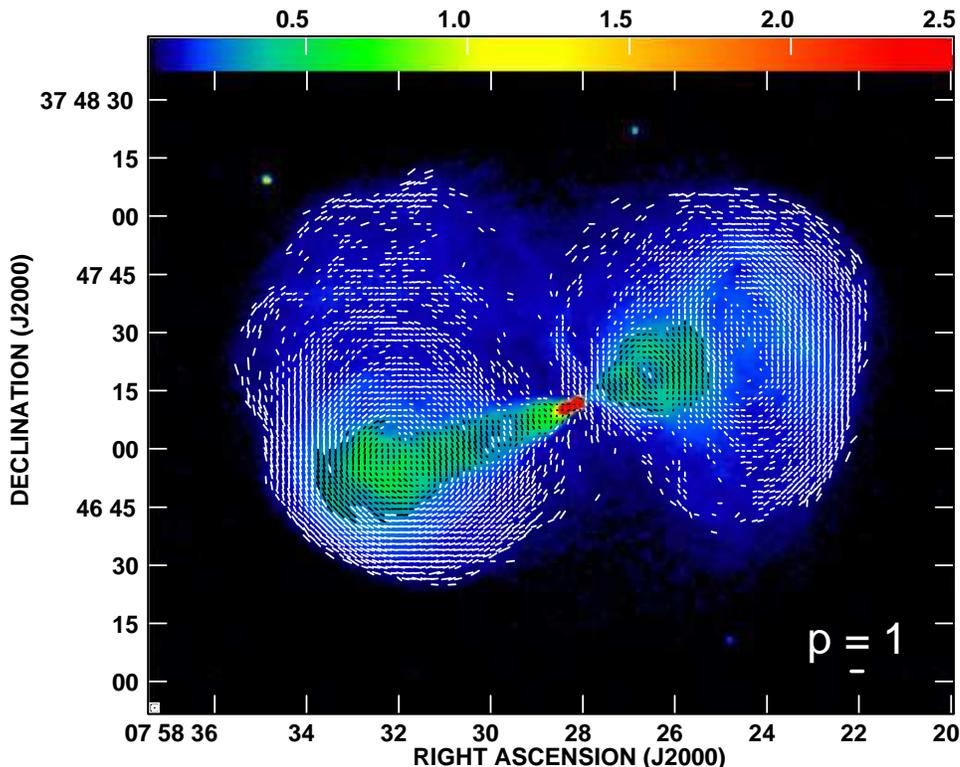}
  \caption{Vectors with lengths proportional to the degree of polarization at
  4.9\,GHz and directions along the apparent magnetic field, superimposed on a
  false-colour display of the total intensity over 0755+37 at 4.9\,GHz.  The
  resolution is 1.3\,arcsec FWHM and the colour-scale range is 0 -- 2.5\,mJy
  beam$^{-1}$.  The vector directions are derived from three-frequency RM fits
  at 4.0-arcsec resolution (Guidetti et al., in preparation).
\label{fig:0755vectors13}}
\end{figure*} 

Figs.~\ref{fig:0755all}(c) and (d) show the intensity gradients over the whole
source at 1.4 GHz, 4.0\,arcsec FWHM resolution and 4.9 GHz, 1.3\,arcsec FWHM
resolution, respectively. These figures 
emphasise the strong differences between the internal structures of the
lobes: multiple recessed ridges with significant brightness gradients in the E
lobe, but much smoother structure in the W lobe away from the
jet. 

Figs.~\ref{fig:0755all}(e) and (f) clearly show three distinct spectral-index
regions on each sides of the source, as follows.
\begin{enumerate}
\item $\alpha \approx 0.6$ at the bases of both jets, in a broad
cap in the NW part of the W lobe, and all along the region delineated
by the strongest brightness gradients in the E jet.
\item $\alpha \approx 0.8$ over most of the rest of both lobes,
including the ridge extending Northward from the nucleus.
\item There is also steeper-spectrum diffuse emission with $\alpha$ increasing from
$\approx$ 1 in the central part of the source to $\approx$1.5 at the N and S
edges.
\end{enumerate}
The spectral-index image suggests that the counter-jet flow persists as far as
the ridge of emission in the W lobe marked on Figs.~\ref{fig:0755all}(a) and (b), despite the
lack of evidence for this in total intensity.

Fig.~\ref{fig:0755all}(g) shows that the apparent magnetic field in both lobes
is exceptionally well organised, and mainly tangential to the lobe
boundaries. The degree of linear polarization is $p\ga 0.6$ over much of both
lobes, consistent with the high degree of organization evident from the
vectors. Note the excellent alignment between the field vectors and the ridges
of high brightness gradient on both sides of the source.
Fig.~\ref{fig:0755vectors13} confirms the exceptional degree of ordering of the
magnetic field in both lobes at 1.3-arcsec FHWM resolution.

\begin{figure*}
\includegraphics[width=17cm]{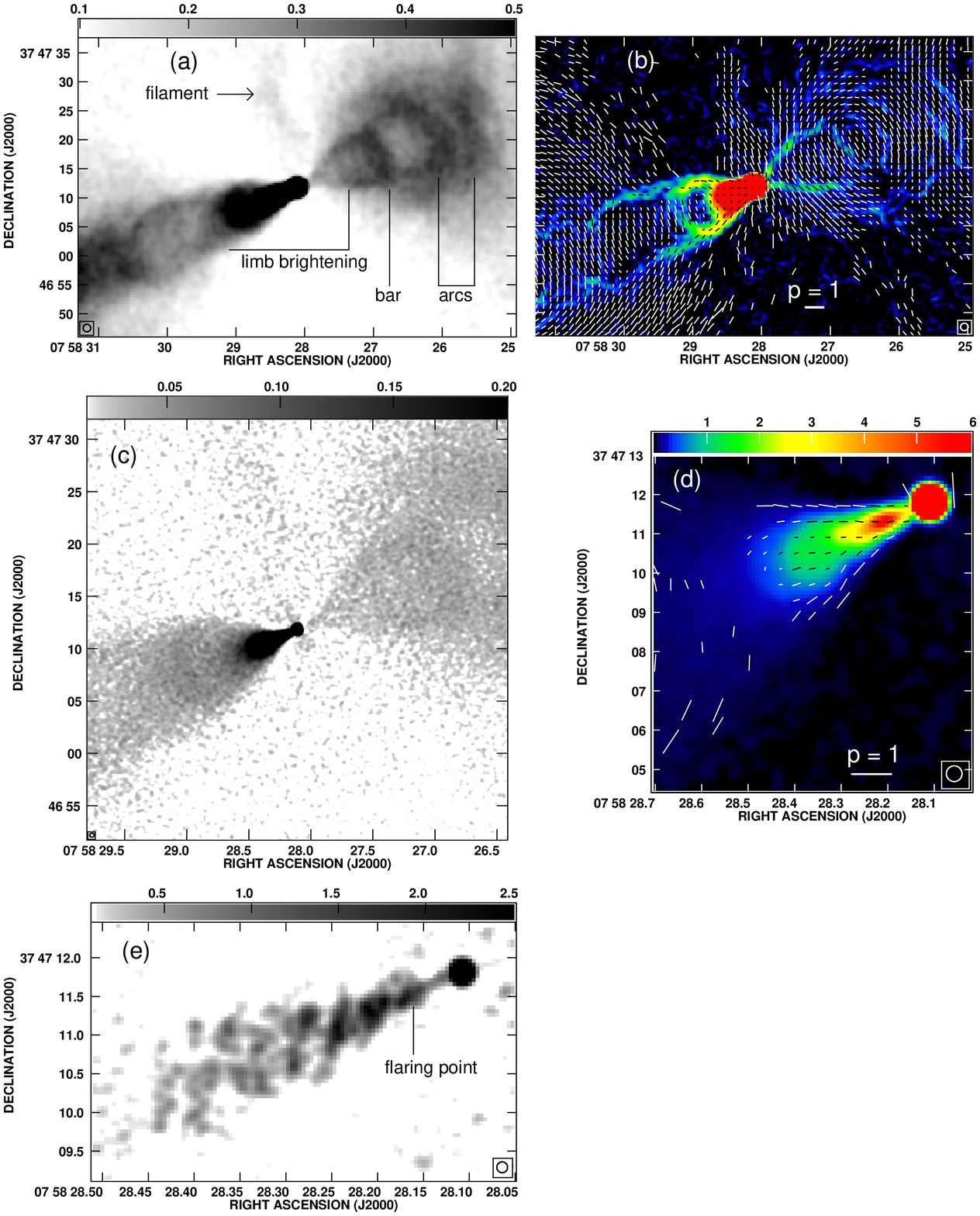}
\caption{High-resolution images of the inner jets of 0755+37. (a) Total
  intensity at 4.9\,GHz with 1.3-arcsec FWHM resolution, plotted with a compressed grey-scale
  range to emphasise fine-scale structure in and around the jets. (b)
  Vectors with lengths proportional to $p_{4.9}$ and directions along the
  apparent magnetic field superimposed on a false-colour image of intensity
  gradient at 4.9\,GHz. The resolution is 1.3\,arcsec FWHM. (c) Total intensity
  at 4.9\,GHz, 0.4-arcsec FWHM. (d) Main jet base at 4.9\,GHz with 0.4-arcsec
  FWHM resolution. ${\bf B}_{\rm a}$ vectors with lengths proportional to $p_{4.9}$ are
  superposed on a false-colour plot of total intensity. (e) MERLIN image of the
  main jet base at 1.7\,GHz with 0.16-arcsec FWHM resolution \citep{Bondi00}. 
  Corrections for Faraday rotation in panels (b) and (d) were made using a 
  three-frequency RM fit at 4.05-arcsec resolution (Guidetti et al., in preparation).
\label{fig:0755hires}}
\end{figure*}   

Fig.~\ref{fig:0755hires} shows the jet bases on larger scales and at higher
resolution. Fig.~\ref{fig:0755hires}(a) is optimised to emphasise fine-scale
structure in the jets at 1.3-arcsec resolution. It shows that the E jet has the
brighter base but becomes limb-brightened at the position indicated on
Fig.~\ref{fig:0755hires}(a) at about 18\,arcsec from the
nucleus. At this resolution, the W counter-jet contains a centre-darkened
structure that expands at about the same rate as the brighter E jet, embedded in
a much broader, and more rapidly-expanding cone of emission with at least two
curved arcs (Fig.~\ref{fig:0755hires}a). The centre of the counter-jet is
crossed by a prominent, straight bar of emission (labelled as such on
Fig.~\ref{fig:0755hires}a) at $\approx$15\,arcsec from the core. The steepest
brightness gradients at the edges of both jets are very similar in form within
$\approx$15\,arcsec of the nucleus so that, as in 0206+35, the {\it inner}
geometry of the counter-jet structure appears to mimic that of the brighter jet
on the other side of the source (Fig.~\ref{fig:0755hires}b).

Fig.~\ref{fig:0755hires}(b) also shows detail of the magnetic field organisation
at the bases of both jets at 1.3-arcsec resolution.  The bright base of the E
jet has the magnetic field roughly parallel to the jet axis, but there is a
rapid transition, with the field becoming perpendicular to the axis as the jet
expands.  On the counter-jet side the field is also transverse. There is little
evidence for any perturbation of the magnetic field structure at the edges of
the jets except at the N edge of the counter-jet where the magnetic field
becomes parallel to the steepest brightness gradient once the jet widens
significantly.  The very high degree of polarization in the surrounding diffuse
emission makes it difficult to disentangle the true polarization of the jets
where their emission is weak, e.g.\ at their edges.

Figs~\ref{fig:0755hires}(a) and (b) also show a filament of faint
emission which extends for about 30 arcsec Northward from the vicinity of the
nuclear source, roughly perpendicular to the jets. It has a very high degree of
linear polarization, with an apparent magnetic field parallel to its length.
Fig.~\ref{fig:0755vectors13} suggests that this highly polarized
filament may be part of a larger region of enhanced polarization that delineates
the inner boundary of the W lobe.

\begin{figure*}
\includegraphics[width=16.5cm]{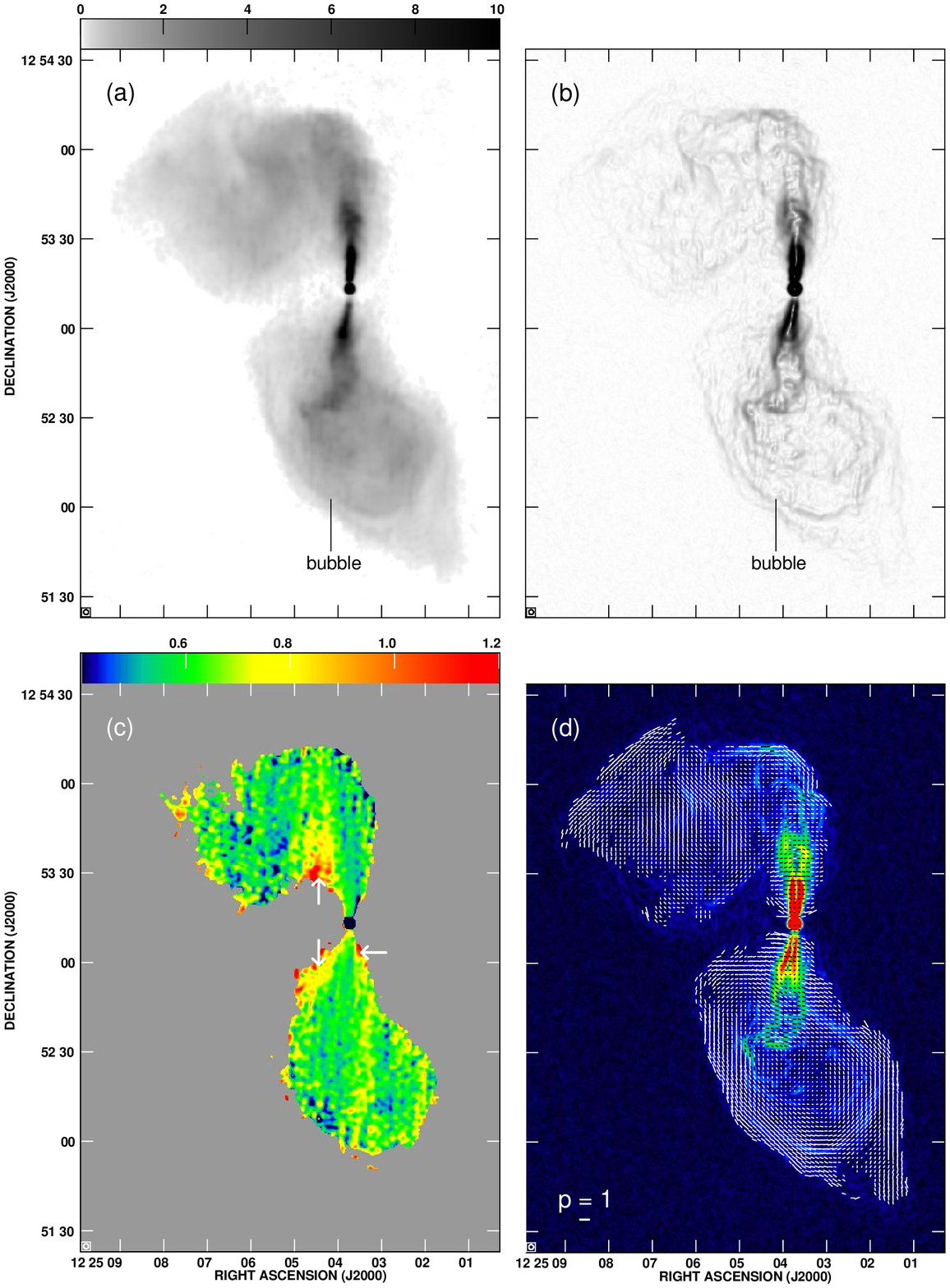}
\caption{Images of M\,84 at 1.65-arcsec FWHM resolution. (a) 4.9-GHz total intensity.
(b) 4.9-GHz intensity gradient.  (c) Spectral index between 1.4 and 4.9\,GHz,
plotted only where its rms error is $<$0.1.  The vertical ``streaks'' are
artefacts. Arrows mark the areas when the spectral index is significantly
steeper than the typical value of $\alpha = 0.6$. (d) ${\bf B}_{\rm a}$ vectors with
lengths proportional to $p_{4.9}$, superposed on a false-colour plot of
intensity gradient. Corrections for Faraday rotation were made using a
4-frequency RM image at 4.5-arcsec resolution \citep{Guidetti11}. All panels
show identical areas.
\label{fig:m84all}}
\end{figure*}

\begin{figure*}
\includegraphics[width=17cm]{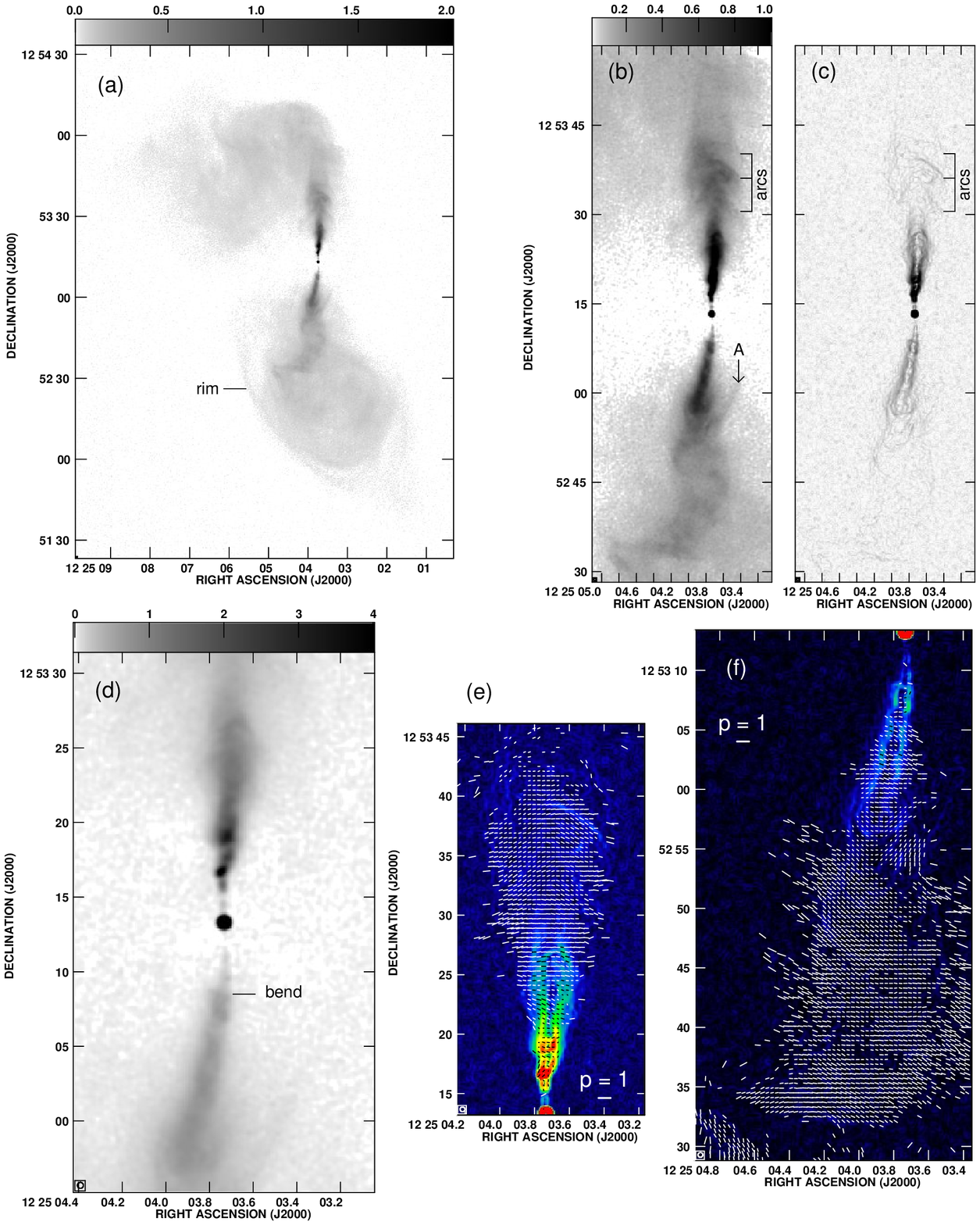}
\caption{4.9-GHz images of M\,84 at 0.4-arcsec FWHM resolution. (a) Total intensity
  for the whole source. (b) Total intensity for the jets. (c) Intensity gradient
  for the jets. Three prominent ``arcs'' in the N jet are labelled on panels (b)
  and (c). (d) Total intensity for the inner jets, showing the abrupt bend in
  the counter-jet. (e) ${\bf B}_{\rm a}$ vectors with lengths proportional to
  $p_{4.9}$, superimposed on a false-colour image of intensity gradient for the
  N jet. (f) As (e), but for the S jet.
\label{fig:m84hires}}
\end{figure*}

Fig.~\ref{fig:0755hires}(c) shows the total-intensity structures at the bases of
the jets at 0.4 arcsec resolution. The main jet is clearly centre-brightened
whereas the counter-jet is not and the brighter edges of the counter-jet lie
mostly outside the region that would be delineated by reflecting the main
jet across the nucleus.  As in 0206+35 (Fig.~\ref{fig:0206hires}a) there is a
narrow collimated structure within which the main jet is systematically brighter
than the counter-jet apparently superposed on a broader structure which is
brighter around the counter-jet than around the main jet. As for 0206+35, we
will show elsewhere (Laing \& Bridle, in preparation) that this structure can be modelled 
as a symmetrical relativistic outflow surrounded by modestly relativistic backflow.
 
The polarization image in Fig.~\ref{fig:0755hires}(d) shows the extent of the
region at the bright base of the main jet in which the magnetic field at the edges
is parallel to the expanding outer isophotes, whereas the on-axis field is
oblique (see also Fig.~\ref{fig:0755hires}b). Finally, a 1.7-GHz MERLIN image of
the main jet base (Fig.~\ref{fig:0755hires}e) shows the position of the flaring
point and the details of the initial expansion.

\subsection{M\,84}
\label{m84}

M\,84 is of particular interest for two reasons: it has a much lower radio
luminosity than the other sources we have studied and it shows very clear
evidence for interaction with the surrounding IGM \citep{M84Chandra}.

Fig.~\ref{fig:m84all}(a) shows the total-intensity distribution over M\,84 at
4.9\,GHz with 1.65-arcsec FHWM resolution\footnote{Lower-resolution (FWHM
  $\approx$4\,arcsec) radio observations, shown by \citet{LB87}, are not
  reproduced here.}.  Both jets of this small (overall extent $\approx$12 kpc),
low-luminosity radio source expand rapidly and deflect within about 1 arcmin.
They are surrounded by diffuse emission (at least in projection) everywhere
except perhaps within a few arcsec of the nucleus.  The initially brighter N jet
can be traced as far as the edge of its lobe, where it bends through
$\approx$90$^\circ$ in projection and decollimates on impact. The bending is
accompanied by strong brightness gradients (Fig.~\ref{fig:m84all}b).  In
contrast, and uniquely amongst the sources in this paper, the S jet (initially
fainter and misaligned with the nucleus) appears to terminate within its lobe
and to feed a bubble-like structure with significant internal brightness
gradients and filaments. The bubble is contained within a smoother, more
elongated structure, at least in projection.  The spectral index between 1.4 and
4.9\,GHz (Fig.~\ref{fig:m84all}c) is constant with $\alpha \approx 0.6$ over the
jets, within the Southern bubble and over most of the N lobe. The only regions
of significantly steeper spectral index that we have detected (marked by arrows
on Fig.~\ref{fig:m84all}c) are on both sides of the south jet base and to the
west of the north jet. The current observations are too noisy to determine the
spectral index in the low-surface-brightness emission outside the southern
bubble. Fig.~\ref{fig:m84all}(d) shows the apparent magnetic field structure
over the whole source at 1.65-arcsec FHWM resolution.  The magnetic field in the
S lobe is broadly circumferential and appears well-aligned with the peak in
brightness gradient at the edge of the bubble.  There is a sudden increase in
the degree of polarization at the edge of the bubble, suggesting a discontinuity
in the field structure at that location. A configuration in which the field is
confined to ellipsoidal shells but is otherwise random (model A of \citealt{L80})
gives qualitatively the correct polarization distribution, but the predicted
variation of $p$ across the lobe is smoother than we observe. The magnetic field
in the N lobe is predominantly perpendicular to the presumed path of the jet
along its mid-line.

Fig.~\ref{fig:m84hires}(a) shows the 4.9-GHz total-intensity distribution over
the whole source at 0.4-arcsec FWHM resolution. This highlights the filamentary
structure in the N lobe, the S bubble and a thin rim of emission around the S
lobe. The intensity and gradient images of the jets at this resolution
(Fig.~\ref{fig:m84hires}b and c) emphasise the edges of both jets and the curved
arcs in the north.  The former also shows a curious thin feature (labelled A)
joining the S jet and the edge of its lobe. The misalignment (non-collinearity)
of the axes of the N and S jets beyond a few arcsec (a few hundred parsecs) from
the nucleus and the initially knotty structure of the N jet, can be seen on a
larger scale in Fig.~\ref{fig:m84hires}(d), which also shows faint emission
close to the nucleus on both sides.

Fig.~\ref{fig:m84hires}(e) and (f) show that, although the apparent magnetic field
in the jets is locally well-organised, there are significant regions where the
field is oblique to the jet axis.  In the outer parts of both jets the field appears to be
predominantly perpendicular to the jet axis, but the jet emission also becomes blended with
that from the lobes.  

M\,84 may be an intermediate case between lobed and tailed sources, showing some
characteristics of each class. The N jet terminates in a sharp bend at the outer
edge of its lobe, as often seen in lobed sources (eg.\ 3C\,296,
Section~\ref{3c296}), but there are hints of a nascent tail structure in the
NE. This is supported by {\em Chandra} imaging of M\,84 \citep{M84Chandra},
which suggests that the NE lobe is breaking out of the surrounding hot
plasma. In contrast, the S jet terminates well within its bubble-like lobe. The
oscillation of the S jet prior to its eventual disruption is very reminiscent of
that of the jets within the S spurs of the tailed sources 3C\,31 and 3C\,449
\citep{3c31ls,KSR}.  The spectral gradients (such as they are) are more
characteristic of lobed sources, with no hint of steepening outwards. The
constancy of the spectral index across the radio structure is not surprising:
given the small ($\approx$12-kpc) size of the source, synchrotron losses are unlikely to
have had enough time to steepen the spectrum at GHz frequencies even in the more
extended regions.
 
\subsection{3C\,296}
\label{3c296}

\begin{figure}
\begin{center}
\includegraphics[width=7.5cm]{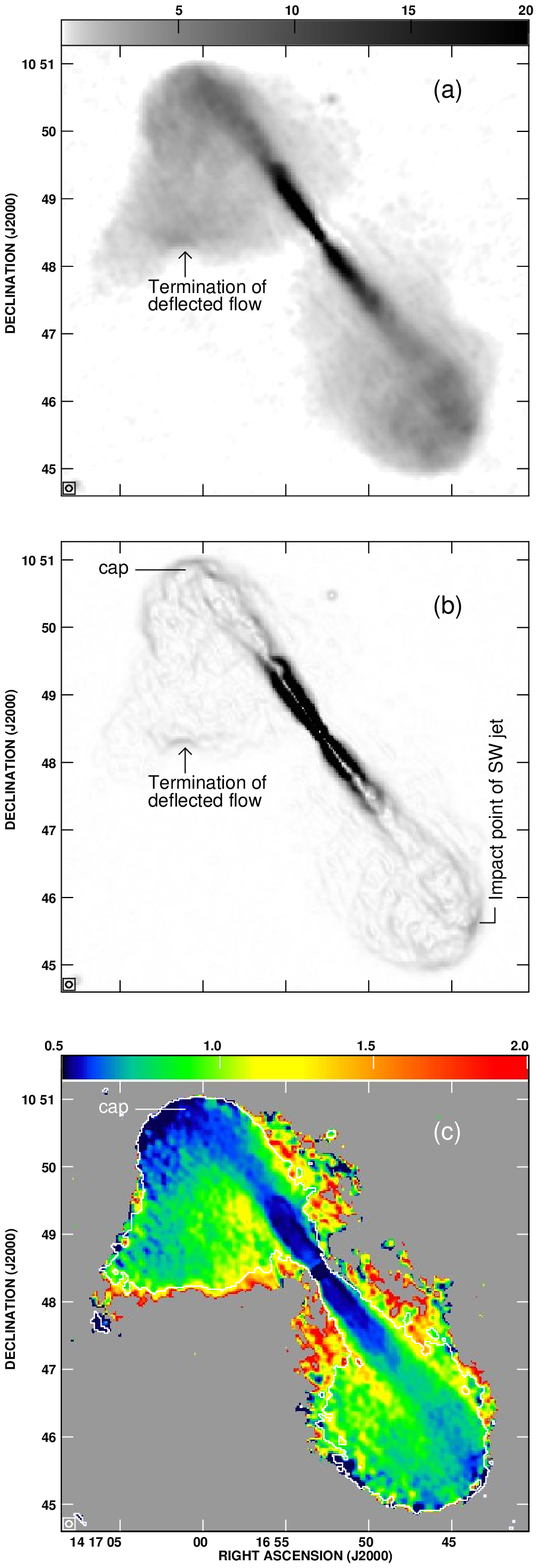}
\caption{(a) Grey-scale of the 1.4-GHz total intensity over 3C\,296 at
5.5-arcsec FWHM resolution.
(b) Intensity gradient image at the same frequency and resolution as panel (a). 
(c) False-colour plot of the 1.4 to 4.9 GHz spectral index distribution at
5.5-arcsec FWHM resolution; data plotted outside the white contour are lower
limits.
\label{fig:296_i_si55}}
\end{center}
\end{figure} 

Fig.~\ref{fig:296_i_si55} shows the 1.4-GHz total intensity and brightness
gradient together with the 1.4 to 4.9-GHz spectral index distributions over
3C\,296 at 5.5-arcsec resolution.  The intensity data are essentially those in
Figs.1(a) and 4(a) of \citet{LCBH06} with an improved deconvolution, but the
grey-scale range in Fig.~\ref{fig:296_i_si55}(a) is chosen to show the jets more
clearly where they appear to enter the lobes.  The corresponding intensity
gradient is shown in Fig.~\ref{fig:296_i_si55}(b). Lower limits to the spectral
indices have been plotted outside the white intensity contour in
Fig.~\ref{fig:296_i_si55}(c) to provide a better representation of the
large-scale spectral gradients at the edges of the radio source.  There is clear
evidence that the flatter-spectrum ($\alpha \approx$ 0.5 to 0.65) jets propagate
to the edges of both lobes, where they deflect and eventually blend with more
extended steeper-spectrum emission whose spectral index $\alpha \approx$ 1. The
NE jet forms a cap of flat-spectrum emission with a semicircular outer boundary, 
again marked by sharp brightness gradients. The flow (as traced by its flatter
spectrum) then turns through $\approx 140^\circ$ in projection, crosses the
entire lobe and impacts on the boundary at the position marked on
Fig.~\ref{fig:296_i_si55}(b). The SW jet, on the other hand, does not form a
cap, but appears to make an oblique impact on the wall of the lobe before
turning through almost 180$^\circ$ in projection back towards the nucleus.  The
spectral index of the more extended emission increases further towards the
centre of the source and towards its outer edges, where $\alpha \approx$ 2.

\subsection{0326+39}
\label{0326}

\begin{figure}
\begin{center}
\includegraphics[width=8.5cm]{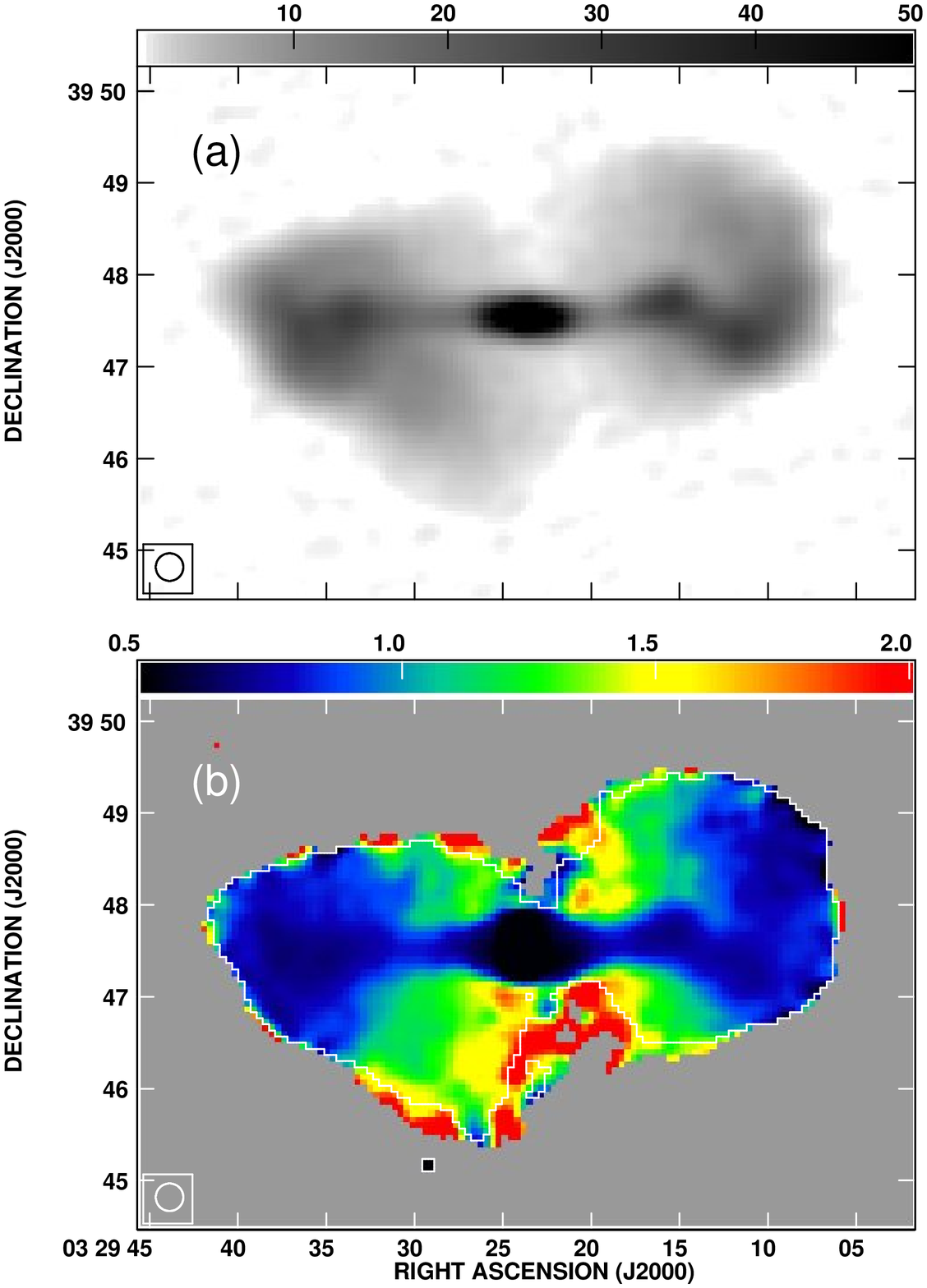}
\caption{(a) Grey-scale of the 1.4-GHz total intensity over 0326+39 at
18-arcsec FWHM resolution.
(b) False-colour plot of the spectral index ($\alpha^{4.9}_{1.4}$) distribution at
the same resolution.
\label{fig:0326_i_si18}}
\end{center}
\end{figure} 

Fig.~\ref{fig:0326_i_si18} shows the distributions of the 1.4-GHz total
intensity and 1.4 to 4.9-GHz spectral index over 0326+39 at 18-arcsec FWHM resolution.

As in the other sources studied here, the lobes of 0326+39 appear to
surround the jets in projection. Even at this relatively low resolution, the
jets are clearly traceable to the outer parts of the lobes in both total
intensity, where they appear to twist and deflect close to the outer edges
of the lobes, and spectral index, where they can be traced as regions of
significantly lower index.  The W jet exhibits a particularly strong kink
towards the S about 2 arcmin ($\approx$ 60 kpc) from the nucleus; this kink is
clearly replicated in the spectral index distribution.

The extended emission of both lobes also shows a well-defined spectral index
gradient, increasing towards the nucleus from $\alpha \approx 1$ near the broad
caps that appear to be dominated by the outer jet emission to a significantly
steeper spectrum with $\alpha \approx 2$ near the centre of the source.  The
spectral index also appears to increase towards $\alpha \approx 2$ in the outer
part of the faint Southward extension of the E lobe.

\section{Discussion}
\label{discuss}

\subsection{Initial jet propagation}
\label{jets}

There appear to be few, if any, morphological differences between the jet base
regions in lobed and tailed FR\,I sources, whose common properties include the
following.
\begin{enumerate}
\item The initial rapid expansion (flaring) and recollimation of the jets is essentially identical
in both types of FR\,I source.
\item Jet bases usually show significant side-to-side asymmetries, although a few very
  symmetrical examples of each type are known. 
\item With the exception of these symmetrical cases, there are further common
  properties, as follows.
  \begin{enumerate}
  \item  One jet in each source exhibits a bright region
  at its base, often with with non-axisymmetric knots and a predominantly
  longitudinal magnetic field.
  \item Jet brightness and polarization asymmetries are correlated: the apparent
  magnetic field on-axis in the brighter jets is initially longitudinal, but switches
  to transverse at larger distances; that in the fainter jets is always transverse.
  \item The jet/counter-jet ratio falls with increasing distance from the
  nucleus.
  \end{enumerate}
\end{enumerate}
These regions at the bases of FR\,I jets are also those in which our models
\citep{LB02a,CL,CLBC,LCBH06} show that the jets decelerate from relativistic to
sub-relativistic velocities. The development of a {\it transverse} velocity
gradient across the jets implies that they decelerate primarily by
boundary-layer entrainment of the external medium. We suggest that this
entrainment occurs primarily in the dense, kpc-scale coronae of hot plasma that
surround the nuclei of twin-jet radio galaxies (probably with an additional
contribution from stellar mass loss) and that the entrainment effectively turns
off on large scales, as it must to avoid further decollimation. {\em Chandra}
observations have revealed coronae of this type in 0206+35, 0755+37
\citep{WBH01}, 0326+39 (Hardcastle, private communication), M\,84
\citep{M84Chandra} and 3C\,296 \citep{Hard05,Croston}; NGC\,193 may have a
similar component \citep{Gia11}. The coronae for which data are available have
central electron densities of $10^5$ to $7 \times 10^5$\,m$^{-3}$, central
pressures of $3 \times 10^{-11}$ to $4 \times 10^{-10}$\,Nm$^{-2}$ and core
radii of 0.3 to 2\,kpc. It seems likely that lobe plasma is excluded from the
immediate vicinity of the nucleus by the high-pressure coronae, so the jets in
both types of source initially propagate in essentially identical environments,
unshielded from the IGM. Further evolution of the jets will depend on their
surroundings: if they propagate through low-density lobe material, then
entrainment will effectively cease and they will recollimate to become almost
cylindrical flows, as seen in the sources described here. Entrainment rates are
also likely to be small in jets without surrounding lobes provided that the
external density is low. For example, there is no sign of any lobe surrounding
the inner jets of NGC\,315, yet it has a very small opening angle after
recollimation \citep{CLBC}. This argues for a negligible entrainment rate at
distances $\ga$35\,kpc, and the external density does indeed fall rapidly on
these scales \citep{Croston}. In contrast, we have argued that entrainment
continues at a lower, but still significant rate after recollimation in 3C\,31,
whose inner jets also have no surrounding lobes. In this source, the opening
angle after recollimation is larger, our models indicate continuing deceleration
and there is a hot-gas component with a large core radius in addition to the
inner corona \citep{LB02b}.

One feature of jet propagation appears to be unique to lobed sources, however.
Our high-resolution data for 0206+35 and 0755+37 show that the apparent
difference in opening angle between the main and counter-jets seen at lower
resolution in these sources is a manifestation of a {\it two-component} jet structure.
In both sources, the main jet and the counter-jet appear to contain both narrow
(well-collimated) and broader features on both sides of the nucleus, but the 
better collimated parts of the main jet are centre-brightened while those in the
counter-jet are centre-darkened.  The broader features at the edges of the
counter-jets are also slightly brighter than those of the main jets.  What {\it
appears} to be poorer collimation of the counter-jet at low resolution is now
seen as a narrow centre-brightened jet opposite a similarly narrow
centre-darkened counter-jet, surrounded by broader emission which is slightly
brighter around the counter-jet than around the main jet.  We will explore
explanations for this ``two-component'' jet structure in terms of relativistic
outflow in the well collimated component surrounded by mildly relativistic
backflow in the broader component in a later paper (Laing \& Bridle, in preparation).

\subsection{Jet termination and lobe structure}
\label{lobes}

With the partial exception of M\,84, to which we return at the end of this section, the sources
described in this paper have lobes similar to those in FR\,II sources
(e.g.\ \citealt{AL87,CPDL,K08}).  They exhibit sharp brightness gradients at their outer edges,
spectral indices that steepen towards the nucleus from the outer lobes and
towards the outer extremities of any lobe, off-axis wings of diffuse emission
near their centres, and generally circumferential magnetic fields.  The only
obvious difference from the lobes of FR\,II sources is the evident lack of hot spots at the ends of the FR\,I
jets. Similar spectral gradients occur in a larger sample of lobed FR\,I sources
observed at lower resolution \citep{Parma99}.  Where the jets dominate the total
intensity, they all have similar spectral indices in the range 0.5 to 0.7.  Even
when the jets are not obviously dominant in intensity alone, the observed spectral
index distributions can trace plausible paths for jets towards the edges of the
lobes. This spectral signature implies that steeper-spectrum lobe material has been displaced by
flatter-spectrum jet material along an extended pathway through the lobes. Note that the
spatial resolution of our data relative to the overall source size is higher than for most
published multifrequency imaging of sources in either FR class. Together
with the ability to trace the jet flow via its flatter spectrum and the use of
intensity gradient images to indicate enhanced compression, our data allow us
to identify a number of new types of structures in the lobes.

The jets often terminate in what we have called ``caps'', with the following properties.
\begin{enumerate}
\item They typically occur at the ends of lobes (one example, 0755+37SE,
  appears recessed, perhaps as a result of projection) and are clearly
  associated with jet termination.
\item They are bounded at their leading edges by smooth outer isophotes
  (approximated by segments of circles) with sharp intensity gradients.
\item They also have inner boundaries, again marked by high intensity
  gradients. 
\item Their emission has a flat spectrum, close to that typical of jets ($\alpha
  \approx 0.6$) after accounting for contamination by steeper-spectrum diffuse
  emission.
\item Four out of five examples are fairly symmetrical with respect to the local jet
  axis. The exception is 3C\,296NE.
\end{enumerate}

There are five clear cases of such ``caps'' out of the ten FR\,I lobes we have studied at
  high resolution: NGC\,193E and W, 0206+35NW, 0755+37SE and
  3C\,296NE, and 0755+37NW may well be similar. 

Other jet terminations show some, but not all, of the same features. In
particular, several show enhanced brightness gradients, but without obvious inner
boundaries. In 0206+35SE, the jet bends away from its initial direction and
creates a sharp brightness gradient where it impacts on the side of the lobe;
the associated emission again has a relatively flat spectrum. Both lobes of
0326+39 probably have similar structures, but the available resolution is not
yet high enough to be sure. In 3C\,296SW and M\,84N, the jets remain straight
until they make oblique impacts on the lobe walls at locations marked by high
intensity gradients after which they bend abruptly. The former case also shows
flatter-spectrum emission at the impact point. 0755+37NW has a weak ring of
emission surrounding a flatter-spectrum region, but this is contained entirely
within the outline of the lobe. It is possible that some of these structures
(especially the last) are caps seen from an unfavourable angle.

We see little convincing evidence for enhancements in $|\nabla I|$ {\em crossing} the
jets close to their termination points, such as might be expected from strong shocks.
This implies that the flow is internally sub- or transonic. Our estimates of 
the on-axis flow velocities after the initial rapid deceleration range from
$\la$0.1$c$ to $\approx$0.6$c$, compared with an internal sound speed of
$3^{-1/2}c = 0.58c$ for an ultrarelativistic plasma. Unless there is significant
deceleration on scales larger than we model, the implication is that the jets
must be very light and energetically dominated by relativistic particles and
magnetic field. 

The lobes in this class of source are very different from the subsonic, buoyant
plumes which are thought to form the outer structures of large FR\,I sources
like 3C\,31 \citep{3c31ls}. The picture that emerges for jet termination in
lobed FR\,I sources is that the flow can be traced at least as far as the end of
the lobe via its flatter spectrum. Where it impacts on the lobe surface, a
high-pressure region (the cap) can be created. The smooth shape of the outer
isophotes (compared with the more ragged outline of the lobes) suggests that the
forward expansion is at least mildly supersonic with respect to the external
medium.  Jet material flows through the cap and back into the lobe, eventually
mixing with pre-existing lobe plasma. The flow pattern is sometimes consistent
with axisymmetry (or at least appears so in projection) but can bend by large
angles without completely losing its collimation.  The clearest example is
3C\,296NE, where the flow bends by $\approx$140$^\circ$ in the plane of the sky
and can be traced as far as the trailing edge of the lobe, where its impact is
marked by a sharp brightness gradient.  An alternative to the formation of a cap
appears to be an oblique collision with the boundary of the lobe. In at least
one case, 0206+35SE, the jet deflects from its original straight path before
hitting the edge of the lobe. This raises the possibility that the impact point
moves around the surface of the lobe, extending it in different directions at
different times and lowering the average advance speed of the lobe compared with
the instantaneous speed of the impact point, as in the ``dentist's drill''
model of FR\,II sources \citep{dentist}.

As noted in Section~\ref{m84}, M\,84 appears to be an intermediate case, in
which only one of the jets appears to impact on its lobe boundary, but no tails
have (yet) developed.  The S lobe of M\,84 is morphologically very similar to
the spurs in 3C\,31, albeit on a much smaller linear scale. This adds to the
developing picture of the transition between fast, well-collimated jets and slow
plumes in tailed FR\,I sources, which is often a two-stage process. The jets
first enter bubbles, within which they disrupt, often thrashing around (as in
M\,84S) before disintegrating completely.  The tails are then formed by escape
of material from the bubbles along the direction of steepest pressure gradient
rather than directly from the jet flow.  A similar morphology is often found in
wide-angle tail sources \citep{HS04}.

\section{Summary}
\label{summary}

We have presented deep, high-resolution, multi-configuration VLA imaging of four
FR\,I radio sources: NGC\,193, 0206+35, 0755+37 and M\,84, together with
lower-resolution observations of 0326+39 and a reanalysis of our published
images of 3C\,296. These sources are all examples of ``lobed'' FR\,I radio
galaxies.  Our results, displayed as images of total intensity, brightness
gradient, degree of polarization, apparent magnetic-field direction and spectral
index, show common features, as follows.
\begin{enumerate}
\item All of the sources have twin radio jets, with side-to-side brightness
  ratios decreasing with distance from the nucleus in a manner qualitatively
  consistent with relativistic, decelerating flow.
\item The brightness and polarization distributions of the inner jets are very
  like those in tailed radio sources, indicating  similar deceleration
  physics. We suggest that the jets in both classes of source 
  propagate unshielded from the surrounding IGM within dense, kpc-scale
  coronae, leading to efficient boundary-layer entrainment.     
\item Farther from the nucleus, the jets in both classes of source recollimate.
  This implies that the entrainment rate is low whether or not they are
  surrounded by lobe plasma.
\item 0206+35 and 0755+37 show evidence for a two-component jet structure in
  which a centre-brightened main jet and centre-darkened counter-jet are
  surrounded by broader features that are somewhat brighter on the counter-jet
  side, suggesting that a central relativistic outflow is surrounded by a slower, 
  but still mildly relativistic backflow. 
\item In all but one case (M\,84S), the jets propagate to the ends of their
  lobes.  Continuing, but less well collimated flow can often be traced in
  spectral index or brightness gradient images.
\item Five or six of the ten jets we have studied at high resolution terminate at the
  ends of their lobes in features we call ``caps'' with smooth outer isophotes,
  sharp inner and outer intensity gradients and relatively flat spectra.
\item An additional three out of ten jet terminations are best described as oblique
  collisions of jets with the outer lobe walls: they also show enhanced outer
  intensity gradients and flat spectra and may be caps seen at unfavourable
  angles. 
\item The lobes resemble those in FR\,II sources, with sharp outer brightness
  gradients, spectral indices which steepen towards the nucleus and
  circumferential apparent magnetic fields.
\item There is little evidence for features in the jet brightness distributions
  which can be identified as strong shocks, either at recollimation or where
  they terminate\footnote{Shocks may occur on smaller scales, at or just
  downstream of the flaring point.}. This implies that the flow is internally
  sub- or trans-sonic on large scales.
\end{enumerate}
We will present quantitative modelling of the inner jets in later papers.

\section*{Acknowledgements}

We thank Anita Richards for a useful perl script.

\end{document}